\documentclass[usenatbib,useAMS]{mn2e}

\usepackage{epsfig}

\newif\ifAMStwofonts

\newcommand{\ion}[2]{\hbox{{#1}\,{\sc {#2}}}}

\title[Dust sputtering in the IGM]{IGM metal enrichment through dust 
                                                          sputtering}

\author[S. Bianchi \& A. Ferrara]{
Simone Bianchi$^{1}$
and
Andrea Ferrara$^{2}$\\
$^1$ Istituto di Radioastronomia / CNR - Sezione di Firenze,  Largo
Enrico Fermi 5, 50125 Firenze, Italy\\
$^2$ SISSA/International School for Advanced Studies, via Beirut 2-4,
34013 Trieste, Italy
}

\date{January 2005}

\pagerange{\pageref{firstpage}--\pageref{lastpage}}
\pubyear{2005}
\begin{document}

\maketitle
\label{firstpage}

\begin{abstract}
We study the motion of dust grains into the Intergalactic Medium (IGM) 
around redshift $z=3$, to test the hypothesis that grains can 
efficiently pollute the gas with metals through sputtering. We use the 
results available in the literature for radiation-driven dust ejection 
from galaxies as initial conditions, and follow the motion onward. Via 
this mechanism, grains are ejected into the IGM with velocities $>100$ 
km s$^{-1}$; as they move supersonically, grains can be efficiently 
eroded by {\em non-thermal} sputtering. However, Coulomb and collisional 
drag forces effectively reduce the charged grain velocity. Up-to-date 
sputtering yields for graphite and silicate (olivine) grains have been 
derived using the code TRIM, for which we provide analytic fits. 
After training our method on a homogeneous density case, we analyze the 
grain motion and sputtering in the IGM density field as derived from a
$\Lambda$CDM cosmological simulation at $z = 3.27$. 
We found that only large ($a \ga 0.1$-$\mu$m) grains can travel up to 
considerable distances (few $\times 100$ kpc physical) before being stopped. 
Resulting metallicities show a well defined trend with overdensity $\delta$. 
The maximum metallicities are reached for $10<\delta<100$ (corresponding to 
systems, in QSO absorption spectra, with $14.5 < \log N(\ion{H}{i}) < 16$). 
However the distribution of sputtered metals is very inhomogeneous, with 
only a small fraction of the IGM volume polluted by dust sputtering
(filling factors of 18 per cent for Si and 6 per cent for C).
For the adopted size distribution, grains are never completely destroyed; 
nevertheless, the extinction and gas photo-electric heating effects due to 
this population of intergalactic grains are well below current detection limits. 
\end{abstract}


\section{Introduction}

In recent years, the possible presence of dust in the IGM has received new
attention. Before the observations of CMB anisotropies 
\citep{DeBernardisNature2000,SpergelApJS2003} strengthened the hypothesis of 
a flat, $\Lambda$-dominated universe, an explanation alternative to
cosmology was proposed for the dimming of distant ($z\approx$0.5) 
Type Ia SNe \citep{RiessAJ1998,PerlmutterApJ1999}: obscuration from
intergalactic dust may have been responsible for the observations, provided
that dust grains had a {\em grey} extinction law \citep[as for elongated
or large grains;][]{AguirreApJL1999,AguirreApJ1999} and a smooth distribution
tracing the lower density intergalactic gas \citep{CroftApJL2000}. Although 
now the cosmologic origin for the dimming is preferred,  dust is still considered 
to be an important component of the extragalactic medium. Besides the most 
commonly studied extinction effects, dust may affect, for instance, the 
thermal balance of the gas, being an effective heating agent in low density
environments 
\citep[through photoelectric heating;][]{NathMNRAS1999,InoueMNRAS2003,InoueMNRAS2004} 
and providing a source for cooling in the denser, higher temperature 
Intracluster Medium \cite[via IR emission;][]{MontierA&A2004}.

Furthermore, dust is studied for its own nature. About half of the
metals in the Milky Way and in other local galaxies is locked up in dust
grains \citep{WhittetBook1992,JamesMNRAS2002}. Though originating in 
stars, metals are also present in the low density IGM, far from production
sites. Metal lines are now routinely found associated to hydrogen in the 
$z=3$ Ly$\alpha$ forest for column densities as low as N(\ion{H}{i}) $<$ 
10$^{14.5}$, corresponding to gas only 10 times more dense than the cosmic 
mean \citep{CowieNature1998,EllisonAJ2000,SchayeApJ2001}. 
The most popular mechanism to explain the metal pollution of the IGM is 
galactic winds from supernovae explosions: metal enriched gas is blown out 
from (proto-)galaxies into an IGM with pristine conditions 
\citep{MadauApJ2001,AguirreApJ2001a,AguirreApJ2001b}.
As an alternative, \citet{FerraraApJ1991} and 
\citet{AguirreApJL2001,AguirreApJ2001a} have proposed that metals could be 
expelled from galaxies as radiation-pressure driven grains. Grains then 
release metals in the IGM through sputtering. Not only  this mechanism can
in principle provide for the same levels of metallicity as the galactic
winds hypothesis, but the attractive feature of this model is that, unlike 
winds, enrichment by dust would not impact the thermal/structural properties 
of the IGM, as no shock waves are involved in the transport. In addition,
metals transported via this process are associated with cool gas, alleviating 
the need to explain how this can be achieved if metals are transported by the 
hot gas produced by SN shocks.

In spite of such interesting features several questions await a quantitative 
answer: how far can dust grains travel in the IGM? What is the dominant 
sputtering mechanism? How much dust is needed to produce the observed 
metallicity?  These are the basic answers we are aiming at obtaining in the 
present work.

A few authors have studied the ejection of dust grains from galaxies
due to the radiation pressure from the galaxy's starlight. References
to these works are given in \S~\ref{se_eje}. Although no estimate
is made about what fraction of the total dust content is ejected, most
works agree that escaping grains will be able to overcome the gravitational
attraction reaching large velocities, $v\ga$100 km s$^{-1}$. When colliding
with gas particles, the kinetic energy of the impact is comparable to the
thermal energy of a gas with temperature $T\approx$10$^6$-K. Thus, {\em
non-thermal} sputtering may be a viable mechanism to erode grains and
deposit their constituents (metals), even in the low temperature (and
density) gas. In this paper, we will assume that a fraction of galactic
dust grains will be able to go beyond the virial radius of a galaxy, with 
velocities and radii as inferred from literature. From this point onward
the grain is slowed down by drag forces: we include both collisional drag
and Coulomb drag (due to the ionised nature of the medium and the 
charge a grain attains when exposed to the metagalactic UV background).
During the grain motion, metals are deposited in the IGM as a result of
sputtering (we consider both thermal and non-thermal processes). Our 
approach is different from that of \citet{AguirreApJL2001,AguirreApJ2001a}, 
as they assume that dust grains can reach the equilibrium point between 
gravitation and radiation pressure and they include thermal sputtering only.

The paper is organised as follows: in \S~\ref{se_eje} we summarise the
literature results on dust ejection from galaxies, which we use as initial
conditions in our computation; \S~\ref{se_physics} describe the physics we
have adopted to describe the grain motion, charging and sputtering. Results
are presented and discussed in \S~\ref{se_homo} for the ideal case of a 
homogeneous density field; the method is then applied to a more realistic 
simulated cosmological density field in \S~\ref{se_cosmo}. In
\S~\ref{se_confro} we will compare our results with previous works.
Finally, we summarise our results in \S~\ref{se_summa}\footnote{Throughout this paper we will
assume a flat universe with total matter, vacuum, and baryonic
densities in units of the critical density of $\Omega_m = 0.3$,
$\Omega_{\Lambda} = 0.7$, and $\Omega_bh^2 = 0.028$, respectively, and a
Hubble constant of $H_0 = 100\,h$ km s$^{-1}$ Mpc$^{-1}$, with $h=0.7$. 
The parameters defining the linear dark matter power spectrum are 
$\sigma_8=0.9$, $n=0.93$, $dn/d \ln k =0$ \citep{SpergelApJS2003}}.

\section{Dust Ejection From Galaxies}
\label{se_eje}

Following the motion of dust grains from their formation sites in a 
galactic disk to the outer reaches of the halo is a complex task. It 
involves the evaluation of: radiation pressure and other forces 
resulting from anisotropic radiation fields \citep{WeingartnerApJ2001b}, 
possibly taking into account the opacity of the dusty disk
\citep{DaviesMNRAS1998}; disk and halo gravity; gas drag (and the grain
charge; see next Section); sputtering rates; magnetic forces; geometry
and physical condition in a multiphase disk and halo gas.

A few authors have studied the dust expulsion, including most (but not
all) of the relevant processes. We summarise here their findings.

\begin{enumerate}
\item Depending on their size and composition, some dust grains can 
escape the halo of the galaxy in a relatively short time (a few hundreds 
Myr). These escaping grains attain terminal velocities of about 100-1000
km s$^{-1}$ 
\citep{FerraraApJ1991,FerraraProc1997,ShustovARep1995,SimonsenA&A1999}).

\item Escaping grains are relatively large, with sizes in the range 0.05-0.2 
$\mu$m. Larger grains are too heavy to escape the gravitational well, while 
smaller grains offer smaller efficiencies to radiation pressure and do not 
travel far from their formation site \citep{ShustovARep1995,DaviesMNRAS1998}. 
Furthermore, smaller and slower grains could be more effectively eroded
by sputtering, as they spend more time in the hot halo environment
\citep{FerraraApJ1991,ShustovARep1995}. Instead, velocities for escaping 
grains do not depend much on the grain size \citep{SimonsenA&A1999}.

\item Being heavier and having smaller radiation pressure efficiencies,
silicate grains reach smaller velocities than graphite grains and may be
underrepresented among the escaping grains
\citep{BarsellaA&A1989,FerraraA&A1990,FerraraApJ1991}.

\end{enumerate}

Unfortunately, none of these works provide a statistic for the properties 
and amount of escaping grains. Apart from a few illustrative cases, here 
we will simply assume that sizes are in the range 0.05$<a$ [$\mu$m] $<$0.2 
and velocities in the range 100$<v$ [km s$^{-1}$] $<$1000, with all values 
of $a$ and $v$ equally represented. Typically, dust grains in a galaxy are 
believed to obey a power law size distribution, $n(a)\propto a^{-3.5}$, and 
to have a ratio 1:1 for the relative proportion of graphite and silicate grains 
\citep*[i.e. the MRN model for Milky Way dust;][]{MathisApJ1977,DraineApJ1984}.
For a given dust mass, the MRN model sets the number of grains for each radius 
bin, with smaller grains being obviously more represented. By adopting a flat
distribution for the sizes of the grains that eventually escape in the IGM
(i.e. with the same number of large and small grains),
we implicitly assume that a good fraction of the dust mass is trapped inside
the galaxy. Even allowing all large grains to be ejected, at maximum only 
about 10 per cent of the dust mass can travel to the IGM. Thus, we use 10
per cent as an upper limit to the fraction of a galaxy's dust mass that can 
be ejected. The limit has been derived for the MRN graphite and silicate 
abundance ratio 1:1, that we adopt also for ejected grains. As literature 
results suggest a lower efficiency for ejection of silicate grains, this 
ratio may constitute an upper limit to the presence of silicate grains in 
the IGM, if the assumption of MRN-like dust inside a galaxy is valid.

Unfortunately, size and material distributions for dust in external
galaxies are quite uncertain, especially at high redshift. In the early
universe, dust can be efficiently produced only by Type II supernovae 
\citep{TodiniMNRAS2001}. For pristine conditions of the parental cloud,
dust is produced in pair-instability supernovae from massive (140-260
$M_\odot$) stellar progenitors: models predict in this case a preferential 
formation of silicate grains \citep{NozawaApJ2003,SchneiderMNRAS2004}
although considerable masses of carbon can be locked up in dust if the 
CO molecule formation is duly taken into account \citep{SchneiderMNRAS2004}.
As the gas metallicity rises above a critical value 
\citep[$\ga10^{-5} Z_\odot$;][]{SchneiderNature2003} the formation of very
massive star is suppressed and dust forms in the ejecta of core-collapse
supernovae: in this case, similar masses of carbon and silicate grains
can form \citep{TodiniMNRAS2001}. For both kind of objects, the
cumulative size distribution over all the materials formed in the
ejecta can be roughly described by a power law; in particular, large grains 
tend to follow the MRN distribution \citep{NozawaApJ2003}. Thus, we believe 
the upper limits derived above for the amount of dust ejected into the IGM 
are reasonable estimates.

\section{Physics of dust grains in the IGM}
\label{se_physics}

In this Section we present the basic physical ingredients we have adopted
to describe the motion, charging and sputtering processes for dust in the 
IGM. We will consider a gas with pristine composition (76 per cent H, 24
per cent He in mass). 
Before considering a cosmological density field, we will show
some illustrative cases for the mean density at $z=3$. For the adopted 
cosmology, 
the mean baryonic density at $z=3$ is $\rho_\mathrm{b}= 
2.3 \times 10^{-29}$ g cm$^{-3}$ (equivalent to $n_\mathrm{H}=1.0 \times 
10^{-5}$ cm$^{-3}$). As for the gas temperature, we have adopted $T=2\times 
10^4$-K, a typical value derived for the low density IGM from QSOs 
absorption lines \citep{SchayeMNRAS2000}.
As we allow for different ionization fractions, six different kind of 
particles may be present in the IGM: neutral and ionised hydrogen 
(\ion{H}{i}, \ion{H}{ii}), neutral, single and double ionised helium 
(\ion{He}{i}, \ion{He}{ii},\ion{He}{iii}) and electrons. 

We will find useful to express the grain velocity $v$ through the atomic 
speed ratio
\[
s_i=\left(\frac{m_i v^2}{2 KT}\right)^{1/2},
\]
where $m_i$ is the mass of each particle constituting the gas. It is 
$s_\mathrm{H}$=$s_\mathrm{He}/2$=$\sqrt{m_\mathrm{H}/m_\mathrm{e}}
s_\mathrm{e}\approx$42.8$\,s_\mathrm{e}$. We saw in \S~\ref{se_eje}
that escaping grains achieve a certain terminal velocity as a balance
between radiation pressure and the other forces opposing the motion.
That velocity will be considered as the initial velocity of the grain 
in the IGM, and we will follow the motion assuming that dust is slowed 
down because of gas drag (\S~\ref{se_drag}). For simplicity, the gas 
drag is the only force included in our study. This is clearly an 
approximation when studying the motion of a grain in a inhomogeneous 
density field (\S~\ref{se_cosmo}). In that case we should also consider 
the path deviations due to gravitational attraction by denser regions; 
if the encountered overdensities are forming stars, radiation pressure 
and other forces resulting from anisotropic radiation fields 
\citep{WeingartnerApJ2001b} should be also included. However, the
approximation is not too crude: at the virial radius of a $z=3$ galaxy 
of {\em median} mass (\S~\ref{se_homo} and \S~\ref{se_cosmo}), the 
gravitational and radiation pressure force (assuming a galactic mass-to-light 
ratio) on a $a$=0.1-$\mu$m grain nearly compensate each other, being 
of order $10^{-23}$ dyne. The drag force due to the IGM gas is of the 
same order (see Fig.~\ref{fi_fdrag}) and dominates as the grain moves 
from its initial position, since the drag force on high velocity grains 
decreases less than the distance squared.

Grains in the IGM are charged, because of collisions with ions and
because of the photoionization due to the metagalactic UV background.
The charge influences both the drag and the sputtering efficiency. The
charging processes we have considered and the typical charges of a grain
in the IGM are described in \S~\ref{se_charge}. We will denote with 
$U$ the potential on the grain surface ($U=Ze/a$, where $Z$ is the grain 
charge in units of the electric elementary positive charge $e$; $Z=694.5 
U[\mathrm{Volt}] a[\mu\mathrm{m}]$). It is helpful to define the 
reduced potential 
\[
\phi=\frac{e U}{KT}.
\]
Charged grains could be deviated from their path by the Lorentz force, 
if moving through a magnetic field. For simplicity we neglect this force
here. If the primordial IGM magnetic fields are low as predicted in 
cosmological simulations \citep[$B\approx 10^{-19}$ G;][]{GnedinApJ2000},
the Larmor radius would be of order $10$ Gpc for the grain properties 
considered in this work (\S~\ref{se_tgc}). Being this much larger than
the typical length travelled by a grain in a Hubble time, the Lorentz
force could indeed be neglected. Unfortunately, it is not clear if
efficient large scale amplification mechanisms are present. Some 
observations seem to indicate intergalactic B-fields as high as 0.1 nG 
\citep[ and references therein]{ValleeNewAR2004}, in which case charged 
grains would preferentially move along field lines. However, the 
inclusion of the Lorentz force is still prevented by the large 
uncertainties on the field direction, strength and structure.

The sputtering process is described in \S~\ref{se_sput}.
Because of collisions with gas particles, atoms may be knocked off
a dust grain and released in the IGM. In a low temperature medium,
the most relevant mechanism is {\em non-thermal} sputtering, in which 
the energy transfered to the grain's atoms is the kinetic energy of 
gas particles impacting at a supersonic grain speed $v$.
Since we consider grains made of graphite and silicate (olivine), 
sputtering could be able to pollute the IGM with carbon, iron,
magnesium, silicon and oxygen.

Finally, \S~\ref{se_imp} shortly describes how the physical prescription 
are implemented in our numerical code.

\subsection{Drag Forces}
\label{se_drag}

A charged dust grain moving through a gas experiences a drag force due 
to the direct collision with the particles in the gas and to Coulomb 
interactions with the ions. Under the hypotheses that the collisions 
between grains and ions are elastic and that the grains are much smaller 
than the mean free path of the particles in the gas, we can follow 
\citet{DraineApJ1979a} and write
\begin{eqnarray}
F_{drag}&=&\lefteqn{2 \pi a^2 K T}  \nonumber \\
&& \left\{ \sum_{i} n_i\left[ G_0(s_i) 
+ z_i^2 \phi^2 \ln (|\Lambda/z_i|)G_2(s_i) \right] \right\},
\label{eq_fdrag}
\end{eqnarray}
where the summation is over the six kind of particles we have considered
for the gas, $z_i$ is the charge of each particle (in units of $e$;
$z_i=-1$ for electrons) and
\[
\Lambda=\frac{3}{2ae\phi}\left(\frac{KT}{\pi n_e}\right)^{1/2}.
\]
The first part of Eqn.~(\ref{eq_fdrag}) is the term due to the collisional 
drag \citep*{BainesMNRAS1965}, ignoring a negligible contribution due
to Coulomb focusing. The second part describes the Coulomb drag 
\citep{SpitzerBook1962}. For $G_0(s)$ and $G_2(s)$, whose exact formulae 
depend on the error function, we use the simple analytical approximations 
derived by \citet{DraineApJ1979a},
\[
G_0(s)\approx \frac{8 s}{3\sqrt{\pi}}
\left(1+\frac{9\pi}{64}s^2\right)^{1/2}, 
\qquad G_2(s)\approx \frac{s}{(\frac{3}{4}\sqrt{\pi}+s^3)},
\]
which provide an accuracy within 1 and 10 per cent, respectively, for $0<s<\infty$.

\begin{figure}
\epsfig{file=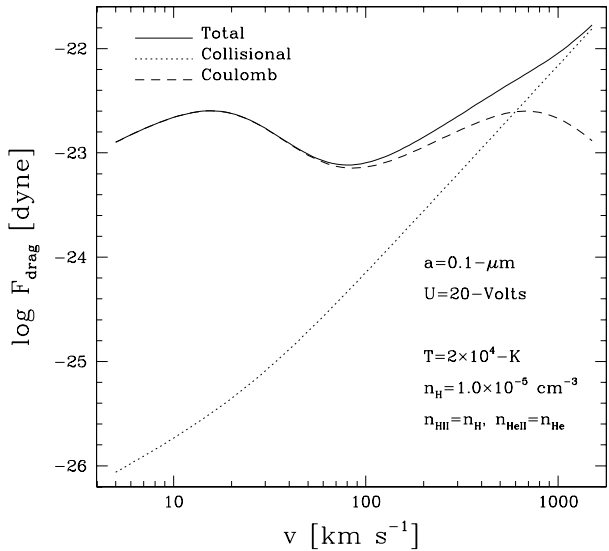,angle=0,width=8cm}
\caption{Drag force on a spherical dust grain of radius $a=0.1$-$\mu$m 
moving at velocity $v$ through a gas with $T=2\times 10^4$-K and 
$z=3$ mean cosmic density. The gas is composed of \ion{H}{ii}, 
\ion{He}{ii} and electrons. The dotted lines is the force component due to 
collisional drag. The dashed line is the component due to Coulomb drag for 
a grain potential $U$=20-Volts. The solid lines represent the total drag.}
\label{fi_fdrag}
\end{figure}

In Fig.~\ref{fi_fdrag} we show the drag force acting on a dust grain 
of radius $a$=0.1-$\mu$m as it moves with velocity $v$ through 
a gas of mean $z=3$ cosmic density and $T=2\times 10^4$-K. We have
assumed that the IGM is composed of \ion{H}{ii}, \ion{He}{ii} and 
electrons. If the grain is not charged (or the gas not ionised) only the 
collisional drag is present (dotted line). For a supersonic dust grain 
($s_i \gg 1$) the collisional term in Eq.~\ref{eq_fdrag} reduces to 
\citep{ShullApJ1978}
\begin{equation}
F_\mathrm{coll}=\pi a^2 v^2 \Sigma_i n_i m_i.
\label{eq_coll_s}
\end{equation}
As it can be seen in Fig.~\ref{fi_fdrag}, the supersonic regime is reached at low 
velocities: for H and He we have $v \gg 18.1$ km s$^{-1}$ and $v \gg 9.0$ 
km s$^{-1}$, respectively (for the gas temperature adopted), and $F_\mathrm{coll} \propto v^2$
beyond that velocity. At high velocities, collisions with helium ions constitute about 30 per cent of the 
collisional drag. Free electrons, instead, do not carry enough momentum to
contribute more than a few percent to the drag force, at any velocity. Thus the collisional 
drag is nearly independent of the ionization state of the IGM.

For a charged grain moving through a plasma, one needs to consider also
Coulomb drag. This is shown in Fig.~\ref{fi_fdrag} 
(dashed line) for a grain with a typical potential $U=$ 20-Volts 
(see \S~\ref{se_tgc}; the Coulomb drag does not depend 
on the sign of the grain charge). For small velocities, the Coulomb drag 
increases linearly with $v$. For supersonic velocities, instead, it decreases 
as $v^{-2}$. The first maximum at $v\approx$ 15 km s$^{-1}$ occurs when the 
dust grain has a velocity similar to the thermal velocity of H and He 
($s_\mathrm{H}\approx 1$ and $s_\mathrm{He}\approx 1$). For larger $v$, 
the contribution of H and He to the Coulomb drag decreases until the electron 
component become dominant. A second local maximum is reached when 
$s_\mathrm{e}=1$ ($v\simeq 680 $ km s$^{-1}$), after which the drag
decreases again. 

It is interesting to note that the Coulomb drag decreases with increasing 
gas temperature. This is because of the increasing number of gas particles 
moving faster than the grain, which do not contribute to the
net drag \citep{NorthropP&SS1990}. The collisional drag, instead, 
depends on $\sqrt T$ for low velocity and is independent of $T$ in 
the supersonic case.

\subsection{Grain Charge}
\label{se_charge}

Dust grains immersed in a hot gas and subject to a UV background attain
an electrical charge because of: i) collisions with electrons (which 
tend to make the charge more negative); ii) collisions with positive
ions, and iii) photoejection of electrons from the grain by absorption of 
UV photons (which tend to make the charge more positive). If we denote with
$J_{\mathrm{c}i}$ the charging rate due to collisions with ion $i$ 
(i.e. the number of charges, in units of $e$, captured by a grain
colliding with particle $i$ per unit time) and with $J_{\mathrm{pe}}$
the rate for photoelectric charging, the grain charge at equilibrium can
be found by imposing
\[
J_\mathrm{pe}+\sum_i J_{\mathrm{c}i} = 0.
\]
We ignore here charge quantization \citep{DraineApJ1987}. This is a safe 
assumption for the large grains adopted in this work and for the large 
potentials they attain (see \S~\ref{se_tgc}).

\subsubsection{Collisional Charging}
\label{se_coc}

We followed \citet{ShullApJ1978} to derive the charging rates on a 
a spherical dust grain that moves relative to the gas. The charging 
rate due to collisions with particle $i$ 
is\footnote{Electrostatic polarization of a dust grain by the
electric field of an approaching charged particle \citep{DraineApJ1987} 
is not included in Eqn.~(\ref{eq_ccharge}). Its effects are small 
for the typical charges described in \S~\ref{se_tgc}.}
\begin{eqnarray}
\lefteqn{J_{\mathrm{c}i}=\pi a^2 n_i (z_i \xi_i + \delta_i)
\left(\frac{KT}{2\pi m_i}\right)^{1/2} \frac{1}{s_i} \times} 
\nonumber \\ 
&&\left\{ 
\vphantom{\left(\frac{KT}{2\pi m_i}\right)}
\sqrt{\pi} \left(\frac{1}{2}+s_i^2-z_i \phi\right)
\left[\mathrm{erf}(s_i-s_o)+\mathrm{erf}(s_i+s_o)\right] \right. 
\label{eq_ccharge} \\
&\lefteqn{+}&\left.
\vphantom{\left(\frac{KT}{2\pi m_i}\right)}
(s_i+s_o)\exp[-(s_i-s_o)^2]+(s_i-s_o)\exp[-(s_i+s_o)^2] 
\right\} \nonumber
\end{eqnarray}
with $\mathrm{erf}$ the error function, $\xi_i$ the sticking 
coefficient of the particle $i$, $\delta_i$ the secondary electron 
emission coefficient and 
\[
s_o = \left\{ \begin{array}{rl} 
0& z_i \phi < 0 \\ 
\sqrt{z_i \phi}& z_i \phi > 0 
\end{array} \right. 
\]
the minimal atomic speed ratio for the approaching particle to win 
electrostatic repulsion. In the limit $s_i\rightarrow 0$, 
Eqn.~(\ref{eq_ccharge}) reduces to the classical result for a 
static grain \citep{SpitzerBook1978}. If the grain moves relative 
to the gas, the frequency of collisions with ions and electrons is altered.
For $s_i\gg1$, Eqn.~(\ref{eq_ccharge}) tends to
\begin{equation}
J_{\mathrm{c}i}=\pi a^2 n_i (z_i \xi_i -\delta_i) v 
(1+\frac{1/2-z_i\phi}{s_i^2}).
\label{eq_cchargesup}
\end{equation}

As for the sticking coefficients, we assume that half of the colliding
electrons are captured by the grain ($\xi_\mathrm{e}=0.5$), the other
half being scattered; and that all the ions will neutralise as they
arrive on the surface, thus sharing their positive charge with the grain
\citep[$\xi_\mathrm{i}=1$;][]{DraineApJS1978,WeingartnerApJS2001}. 
For the coefficients for the secondary emission of electrons, $\delta_i$, 
we used the empirical expressions given by \citet{DraineApJ1979a}. The
coefficients depend on the energy of the impact, $\langle E_0 \rangle +
m_i v^2/2$, where $\langle E_0 \rangle$ is the mean thermal energy of 
the impinging particle \citep{DraineApJ1979a}. Because of the small 
sputtering rates (\S~\ref{se_sput}), we neglect the charge taken
away from the grain by sputtered atoms \citep{DraineApJ1979a}.

\subsubsection{Photoelectric Charging}
\label{se_phc}

The photoelectric charging rate can be written as
\begin{equation}
J_\mathrm{pe}=\pi a^2 \int_{\nu_\mathrm{pet}}^{\nu_\mathrm{max}}
Q_\mathrm{abs}(h\nu,a) Y(h\nu,a) \frac{4\pi J_\nu}{h\nu} d\nu,
\label{eq_photrate}
\end{equation}
where $J_\nu$ is the mean specific intensity of the UV background, 
$Q_\mathrm{abs}(h\nu,a)$ is the absorption efficiency of a dust grain
of radius $a$ and $Y(h\nu,a)$ is the photoelectric yield, i.e. the
probability that an electron is ejected when a photon absorption
occurs. The lower limit of integration is the photoelectric threshold
frequency, $\nu_\mathrm{pet}$, for which we use 
\[
\nu_\mathrm{pet}=\left\{
\begin{array}{ll}
W+eU & U\ge 0, \\
W    & U <  0, 
\end{array}
\right.
\]
$W$ being the workfunction, i.e. the ionization potential of a neutral
bulk material. Thus, if the grain has negative charge, an electron is
photoejected as soon as a photon with energy $h\nu > W$ is absorbed. If
the grain has a positive charge, the photoejected electron needs to have
a kinetic energy large enough to escape the attraction of the charged 
grain, therefore ionizing photons must have $h\nu > W+eU$. 
\citet{WeingartnerApJS2001} derive the photoelectric threshold for spherical
grains taking into account geometric effects and polarization. They
obtain values for $\nu_\mathrm{pet}$ that differ from those used here by 
additional terms of order $e^2/a$ and dependent on $a^{-2}$. For the large 
grains used in this work, those terms are negligible compared to $W$ and 
we omit them here. We use $W=8$-eV and $W=4.4$-eV for graphite and silicates,
respectively \citep{WeingartnerApJS2001}.

Following \citet{WeingartnerApJS2001}, we write the photoelectric yield as
\[
Y(h\nu,Z,a)=y_2(h\nu,a,Z) \min[y_0(\Theta) y_1(h\nu,a),1],
\]
with $y_0$ the photoelectric yield of the bulk material and
$\Theta=h\nu-W$.
For graphite and silicates we use, respectively
\[
y_0^\mathrm{gra}(\Theta)=\frac{0.009(\Theta/W)^5}{1+0.037(\Theta/W)^5},
\qquad
y_0^\mathrm{sil}(\Theta)=\frac{0.5(\Theta/W)}{1+5 (\Theta/W)}.
\]
In small particles, the photoelectric yield is enhanced with respect to
bulk materials. For spherical grains of radius $a$, the enhancement
factor is well approximated by  
\[
y_1(h\nu,a)=\left(\frac{\beta}{\alpha}\right)^2 
\frac{\alpha^2 -2\alpha+2-2e^{-\alpha}}{\beta^2 -2\beta+2-2e^{-\beta}}
\]
with $\beta=a/l_a$ and $\alpha=a/l_a+a/l_e$, $l_a$ and $l_e$ being
the photon attenuation length  and the electron escape length 
\citet{DraineApJS1978}. We adopted $l_e=10$\AA\ \citep{WeingartnerApJS2001}. 
Using tabulated values for the optical properties of graphite and 'smoothed 
astronomical silicates'\footnote{Available at {\tt 
http://www.astro.princeton.edu/\~{}draine}.}
\citep{WeingartnerApJ2001a}, we derived the photon attenuation length with 
$l_a = \lambda/[4\pi\mathrm{Im}(m)]$, where $\lambda$ is the radiation
wavelength and $m(\lambda)$ the complex refractive index. A proper weighted 
mean has been used to take into account the anisotropy of graphite 
\citep{WeingartnerApJS2001}.

Laboratory measures on bulk materials have shown that the distribution
of kinetic energy for photoejected electrons drop to zero at $E=0$ and
$E=h\nu-W$ and peaks at intermediate energies \citep{DraineApJS1978}.
\cite{WeingartnerApJS2001} adopted a parabolic energy distribution and
derived the probability for electron escape to infinity, $y_2$. Again
ignoring all terms of order $e^2/a$ and dependent on $a^{-2}$, it is
\[
y_2(h\nu,a,Z)=\left\{
\begin{array}{ll}
(1-\frac{eU}{h\nu-W})^2 (1+\frac{2eU}{h\nu-W})& U \ge 0,  \\
1 & U < 0. 
\end{array}
\right.
\]

Finally, we have used tabulated values of $Q_\mathrm{abs}(h\nu,a)$ 
for spherical grains of graphite and 'smoothed astronomical silicates'
\citep{WeingartnerApJ2001a}. We did not consider photodetachment of 
electrons in the energy levels above the valence band of negatively charged 
grains \citep{WeingartnerApJ2001a}, which can be ignored for the grain
charges and radii considered here.

\begin{figure}
\epsfig{file=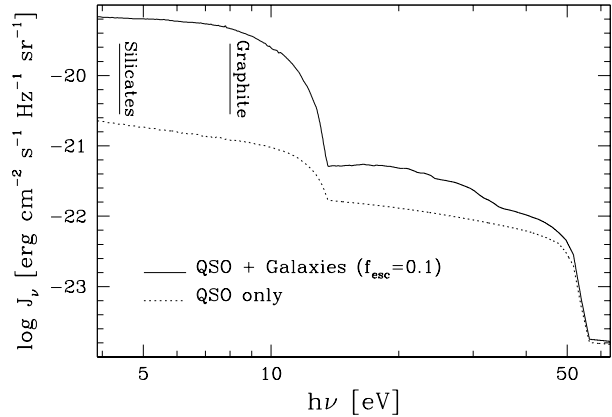,angle=0,width=8cm}
\caption{The z=3 UV background of \citet{BianchiA&A2001} including the
contribution of galaxies and QSOs (solid line). The vertical bars show
the workfunction adopted for silicates and graphite. The dotted line 
is the QSOs contribution to the background.
}
\label{fi_uvbg}
\end{figure}

\subsubsection{Typical grain charges in the IGM}
\label{se_tgc}

\begin{figure*}
\epsfig{file=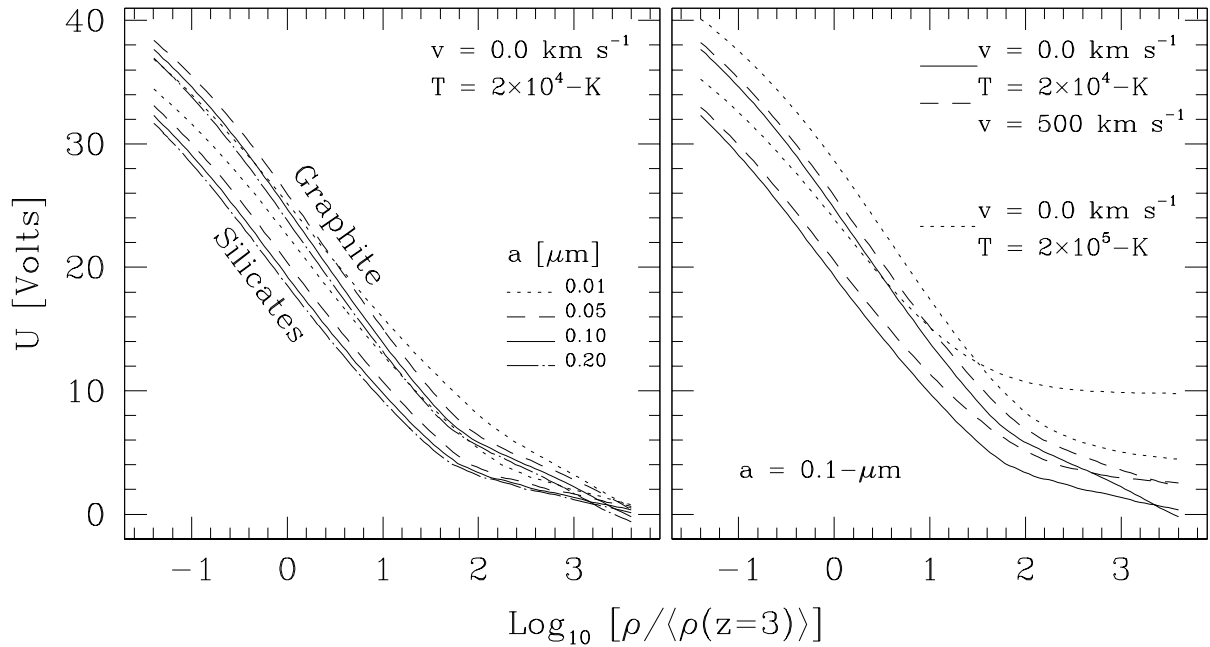,angle=0,width=16cm}
\caption{Equilibrium potentials for graphite and silicate dust grains 
exposed to the $z=3$ UVB, as a function of the gas overdensity. In the
left panel we show the equilibrium potentials for static ($v=0$) grains 
of radii 0.01, 0.05, 0.1 and 0.2$-\mu\mathrm{m}$ (in both panels, the upper bundle of curves 
refers to graphite grains, the lower to silicates). We have assumed a 
gas made of \ion{H}{ii}, \ion{He}{ii} and electrons, with 
temperature $T=2 \times 10^4$-K. In the right panel we show the effects of 
variation in grain velocity and gas temperature for a grain with radius 
$a$=0.1-$\mu$m. For each grain material, the solid line refer to the same 
conditions as in the left panel, the dashed line to a grain moving with 
$v=500$ km s$^{-1}$ in a $T=2 \times 10^4$-K gas, the dotted line to a 
static grain in a $T=2 \times 10^5$-K gas.}
\label{fi_charges}
\end{figure*}

We adopt in this work the ultraviolet background (UVB) of 
\citet*{BianchiA&A2001}. The UVB, shown in Fig.~\ref{fi_uvbg},
includes contributions from both QSOs and galaxies, assuming that a 
fraction $f_\mathrm{esc}=0.1$  of the \ion{H}{i}-ionising UV 
light can escape internal absorption. For any observer's redshift, the 
UVB shows two breaks at frequencies $\nu>13.6$-eV and 
$\nu>54.4$-eV in the observer's rest frame, due to the absorptions of
\ion{H}{i} and \ion{He}{ii} Lyman continuum photons, respectively, from
the residual neutral gas in the IGM. While the QSO contribution (dotted
line) can be fitted between the breaks by a power law (mainly because 
of the assumption on the intrinsic QSO's spectrum), the \ion{H}{i}-ionizing 
UVB including galaxies is only roughly described by $J_\nu\propto\nu^{-1.9}$ 
for $\nu<54.4$-eV. The contribution to the photoelectric 
charging rates of frequencies $\nu>54.4$-eV is negligible, therefore we use
$\nu_\mathrm{max}=54.4$-eV in Eqn.~(\ref{eq_photrate}). The UVB is
computed integrating light coming from all objects up to a maximum redshift 
$z_\mathrm{max}$. In \citet{BianchiA&A2001} we adopted $z_\mathrm{max}=5$.
This assumption has no effect on \ion{H}{i}-ionising photons: because of 
IGM absorption only local sources contribute to the calculation. For
radiation at lower frequencies not able to ionise \ion{H}{i}, the UVB
depends on the redshift assumed for the first sources, mainly because 
of the nearly constant galaxy emissivity at $z>2$ \citep{BianchiA&A2001}. 
However, as long as we use the UVB at $z<4$, the photoemission rate in 
Eqn.~(\ref{eq_photrate}) is only slightly affected by our choice for 
$z_\mathrm{max}$.

In Fig.~\ref{fi_charges} we show the equilibrium potentials for graphite
and silicate grains exposed to the $z=3$ UVB, as a function of the gas
overdensity. Again, the gas is composed of \ion{H}{ii},
\ion{He}{ii} and electrons and has $T=2 \times 10^4$-K. In the left panel
equilibrium potentials are shown for static ($v=0$) grains of radii 
0.01, 0.05, 0.1 and 0.2-$\mu$m (spanning the range of
radii relevant to this work). Because of the different photoelectric
yields and absorption efficiencies, graphite grains generally attain a 
larger (more positive) potential than silicate grains.

In the low density IGM the grain potential reach a relatively high
positive value (a few tens of Volts), in agreement with what found by \citet{NathMNRAS1999}. 
The equilibrium potential is reached when the collisional charging rate
for electrons $J_\mathrm{ce}$ is balanced by the photoelectric charging
rate $J_\mathrm{pe}$, the contribution of positively charged ions being
greatly reduced by Coulomb repulsion ($J_\mathrm{ci} \ll J_\mathrm{pe}$). 
While the collisional charging rate depends on the particle density, the 
photoelectric charging does not. Thus, for low values of $n_e$, a high 
Coulomb attraction is needed to increase the collisional cross section, in 
order to have $J_\mathrm{ce} \approx J_\mathrm{pe}$. 
Charges for the grains studied here are always much lower than the 
maximum limit over which field emission of positive ions occurs 
\citep{DraineApJ1979a}.

At higher density, the contribution of collisions to charging becomes 
progressively more important and the charge decreases.
This partly explains why the slope of the equilibrium potential of a
grain of given radius changes at higher densities (see Fig.~\ref{fi_charges}).
A second reason for this is that $\nu_\mathrm{pet}$ goes across the
Lyman discontinuity in the UVB. For very high densities, only 
collisional charging rates are important. Neglecting secondary emission,
the classical result is recovered: grains have negative 
potential because of the higher collision rate of electron in a gas 
\citep{SpitzerBook1978,OsterbrockBook1989}. For the plasma composition 
and sticking coefficients adopted here, the equilibrium would be reached 
for $\phi\approx -2.0$.

For a grain that moves at a supersonic speed, the collisional charging
rate tends to Eqn.~(\ref{eq_cchargesup}) and becomes larger than that in
the static case. When $s_i \gg 1$ but still $s_e \ll 1$, 
$J_\mathrm{ci}$ gives a constant contribution to the positive charging rate,
with a small dependence on the grain potential. When also $s_e \gg 1$,
$J_\mathrm{ce}$ and $J_\mathrm{ci}$ almost compensate, apart from the
contribution of gas particles impacting from directions normal to the
grain motion, which results in a net negative charging rate proportional 
to $\phi/s_e^2$. Because of this, moving grains always have higher 
(more positive) equilibrium potentials. In Fig.~\ref{fi_charges} 
(right panel) we show the equilibrium potentials for a $a$=0.1-$\mu$m 
grain moving in a $T=2\times 10^4$-K gas at $v=500~\mbox{km s}^{-1}$. 
For the velocities explored in this work, differences with the static 
case are never larger than about 5-Volts. 

In Fig.~\ref{fi_charges} (right panel) we also show the change in the
equilibrium potentials of a $a$=0.1-$\mu$m static grain when the 
temperature raises to $T=2\times 10^5$-K gas. At higher temperatures gas
particles move faster and the Coulomb cross section reduces. At low
densities, a reduced $J_\mathrm{ce}$ has to compete with a $J_\mathrm{pe}$ 
unaltered by temperature changes and the equilibrium potential increases.  
According to the classical result, at high densities the charge would reach 
a high negative value. This is not the case when the secondary emission of 
electrons is included. The secondary emission of electrons decreases the 
negative charging rate due to electrons $J_\mathrm{ce}$ and increases the 
positive charging rates for ions $J_\mathrm{ci}$. As a results, equilibrium 
potentials are higher when this process is included. The effect can be larger for 
higher temperatures (or for supersonic grains). In Fig.~\ref{fi_charges}, 
the positive equilibrium potential attained in the high density gas for 
$T=2\times 10^5$-K is due to secondary emission. 
If $\delta_\mathrm{e},\delta_i$=0, a static grain would have an 
equilibrium potential $U=-34.5$-Volts, while it is 4.2 and 9.8-Volts for 
graphite and silicates, respectively, when the secondary emission is 
accounted for (silicates have higher secondary emission for impinging 
electrons). Thus, secondary emission prevents dust grains in a hot overdense 
gas to reach high negative charges. 
The limiting negative potential below which field emission of electrons
occurs \citep{WeingartnerApJS2001} is never reached in this work.
In the $T=2\times 10^4$-K gas, secondary emission raises the equilibrium
potential by less then a few Volts, and its effect can be mimicked by 
reducing the electron sticking coefficient to $\xi_\mathrm{e} \approx 0.4$ 
(in most cases, the secondary emission due to colliding electrons is more 
important than that due to positively charged ions). 

As the parameters describing the galactic contribution to the UVB (namely, the
star formation history and the $f_\mathrm{esc}$ fraction) are quite uncertain,
especially at high-$z$, we also computed charges when only the QSO contribution 
(dotted line in Fig.~\ref{fi_uvbg}) is taken into account. Because of the reduced
UV flux, the equilibrium charges are smaller in this case. However, the difference 
is small (a couple of Volts) and it does not modify significantly the results
we present in the rest of the paper.

\begin{figure*}
\epsfig{file=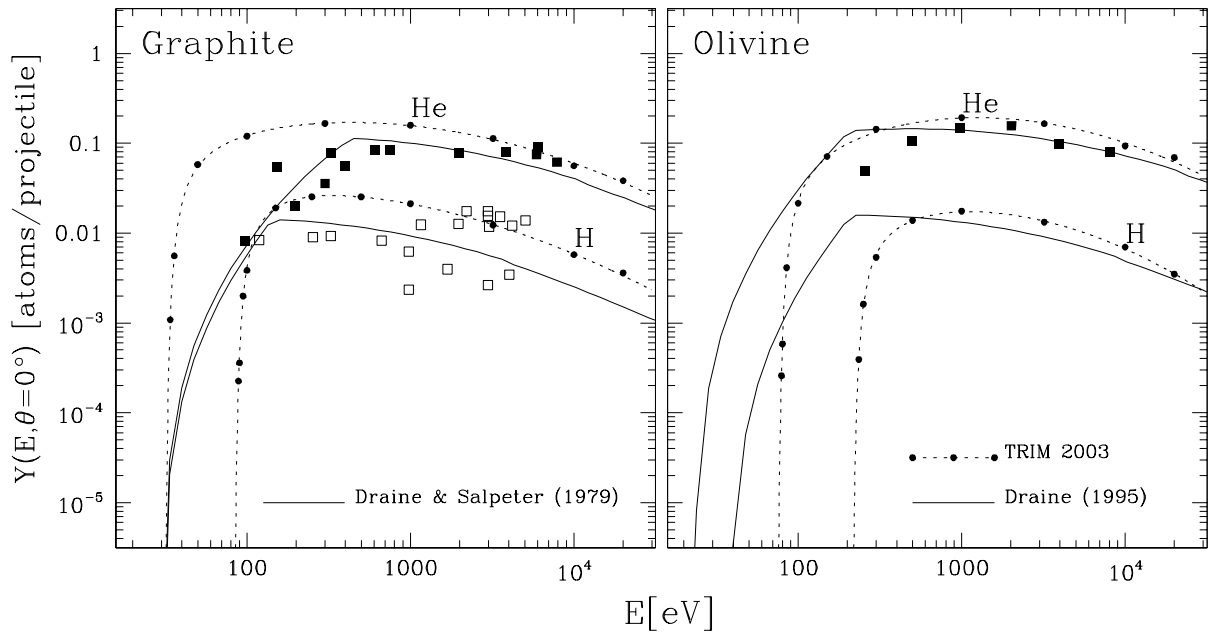,angle=0,width=16cm}
\caption{Normal incidence ($\theta=0^\circ$) sputtering yields for 
H and He on graphite and silicates (olivine), as a function of the 
impact energy. Solid curves are theoretical estimates from 
\citet[][graphite]{DraineApJ1979a} and \citet[][olivine]{DraineApSS1995}.
Small dots are our results using the TRIM 2003 code, together with
the fits of Eq.~(\ref{eq_yfit}) (dotted lines; see text for details). 
Squares are experimental measures from \citet{DraineApSS1995}: closed 
symbols are for He, open symbols for H. As there are no measures for 
sputtering on silicates, datapoints in the right panel are for a similar 
material, Si O$_2$ \citep{DraineApSS1995}.
}
\label{fi_yield0}
\end{figure*}

\subsection{Grain Sputtering}
\label{se_sput}

The efficiency of the sputtering process is given by the sputtering
yield, $Y_i(E,\theta)$, i.e. the number of atoms or molecules of the 
target material that are sputtered in each collision with a projectile
atom of material $i$. For each target/projectile combination, the 
sputtering yield depends on the impact energy $E$ and on the angle 
$\theta$ (relative to the surface normal; $\theta=0$ for normal impact) 
of the impacting particle. Our targets are dust grains made of graphite
or silicates, while we will consider neutral and ionised hydrogen and 
helium atoms as projectiles. The sputtering yield does not depend explicitly
on the charge of the projectile (ions rapidly neutralise when
approaching a solid surface; \S~\ref{se_coc}), although the energy 
of the impact $E$ depends on Coulomb repulsion/attraction. 

From $Y_i$ we can derive the sputtering rate, i.e. the number of atoms that 
are sputtered off each grain per unit time, $dN/dt$. For a charged 
(non-rotating) spherical dust grain of radius $a$ that moves with velocity $v$ 
through a Maxwellian gas, we can write \citep{DraineApJ1979a}
\begin{equation}
\frac{dN}{dt}=\pi a^2  v \sum_i n_i \int_{\epsilon_\mathrm{min}}^{\infty}
g_i(\epsilon,\phi) \langle Y_i [E=(\epsilon-z_i\phi)KT] \rangle_\theta d\epsilon,
\label{eq_fsputrate}
\end{equation}
where again the summation is over all relevant projectile particles of
species $i$, $\epsilon_\mathrm{min}=\max[0,z_i \phi]$ and 
$\langle Y_i (E) \rangle_\theta$ is the angle averaged sputtering yield, given by
\begin{equation}
\langle Y_i (E) \rangle_\theta=2 \int_{0}^{\pi/2} Y_i(E,\theta)
\sin\theta\cos\theta d\theta.
\label{eq_meany}
\end{equation}
The function $g_i$ is given by
\[
g_i(\epsilon,\phi)= \frac{ \exp\left( -s_i^2-\epsilon\right) }{s_i^2} 
\sqrt{\frac{\epsilon}{\pi}} 
\left(1-\frac{z_i\phi}{\epsilon} \right) \sinh (2\sqrt{\epsilon} s_i).
\]
For supersonic grains ($s_i\rightarrow\infty$) the function $g_i$ tends
to
\[
g_i(\epsilon,\phi)= \left(1-\frac{z_i\phi}{\epsilon} \right)
\delta(\epsilon - s_i^2)
\] 
and the rate for {\em non-thermal} sputtering is obtained,
\begin{equation}
\frac{dN}{dt}=\pi a^2 v \sum_i n_i \left(1- \frac{2 z_i e U}{m_i v^2}\right)
\langle Y_i [\frac{1}{2}m_i v^2 - z_i eU] \rangle_\theta,
\label{eq_asputrate}
\end{equation}
where the term within round brackets is a correction to the grain cross
section due to Coulomb focusing \citep{SpitzerBook1978}.

A few analytical models for the sputtering yield of astronomical dust
materials can be found in the literature \citep{DraineApJ1979a,TielensApJ1994}.
As a comparison, we refer here to the sputtering yields of \citet{DraineApJ1979a} for graphite and
for olivine \citep[a typical silicate, FeMgSiO$_4$;][]{DraineApSS1995}.
Normal incidence sputtering yields are shown in Fig.~\ref{fi_yield0}, together 
with a few experimental results which are taken from \citet{DraineApSS1995}. 
Above a certain threshold energy, the yield increases steeply, reach a maximum 
and then decreases slowly for higher impact energies. The normalization of 
$Y_i(E,\theta=0^\circ)$ was chosen to fit the available data. Because of the 
lack of experimental data for low impact energies, which is the most 
important in astrophysical applications, the sputtering yield in proximity 
of the threshold is poorly constrained.
The dependence on the collision angle is uncertain as well, and it is usually 
assumed that $Y_i(E,\theta)\sim 1/\cos\theta$. Under this assumption, it 
follows from Eqn.~(\ref{eq_meany}) that the angle averaged sputtering yield 
is twice the normal one.

\citet{FieldMNRAS1997} and \citeauthor{MayMNRAS2000} (\citeyear{MayMNRAS2000}; 
see also \citealt{FlowerMNRAS1996,JuracApJ1998}) derived sputtering yields 
using the Monte Carlo code for the TRansport of Ions in Matter 
\citep[TRIM; ][]{ZieglerBook1985}. Briefly, TRIM simulates the bombardment 
of a plane-parallel target by projectiles of given kinetic energy and 
impact angle with respect to the surface. The target is simulated as
amorphous (i.e. given the density $\rho$, the position of each atom 
inside the target is random). The depth inside the material at which the 
projectile hits a target atom is derived from interatomic potentials dependent 
on the colliding elements. 
If an atom in the target is given an energy larger than the Displacement Energy 
$E_D$, it will be able to escape from its position (the lattice site) and
it will lose to the lattice an amount of energy given by $E_B$, the Bulk
Binding Energy. Recoiling atoms may collide themselves with other 
target atoms. TRIM follows the whole collision cascade. Finally, an
atom in the proximity of the target surface is counted as sputtered if
the component of its kinetic energy normal to the surface is larger
than the Surface Binding Energy, $E_S$. Several projectiles are needed to 
derive a mean sputtering yield. In order to have an estimate alternative 
to the widely used \citet{DraineApJ1979a} yields, we have used the 2003 
version of TRIM\footnote{The TRIM program is part of 
the SRIM package, which can be downloaded at {\tt http://www.srim.org}.}
to derive the sputtering yield for graphite \citep[$\rho$=2.266 g cm$^{-3}$, 
$E_D$ = 25 eV, $E_B$ = 3.0 eV, $E_S$ = 7.41 eV;][]{FieldMNRAS1997} and 
olivine \citep[$\rho$=3.843 g cm$^{-3}$, $E_D$ = 50 eV, $E_B$ = 9.7 eV, 
$E_S$ = 5.64 eV;][]{MayMNRAS2000}. 

We ran TRIM to have $Y_i(E,\theta)$ for several values of $E$ and
$\theta$. Our results for sputtering of olivine by He are very similar
to those published by \citet{MayMNRAS2000}. The sputtering yields 
we obtain for He into graphite are at least a factor of two larger than
those reported by \citet{FieldMNRAS1997}, probably because they used 
an earlier version of the code. None of those author gives sputtering 
yields with H as a projectile. In Fig.~\ref{fi_yield0} we show the
sputtering yields for $\theta=0^\circ$. Above a certain threshold
energy, yields increase steeply, reach a maximum and decrease at a
slower rate for higher energies (because projectiles implant deeper 
into the target). At high energies, the behavior of the TRIM results
is similar to the analytical models of \citet{DraineApJ1979a}, although 
graphite yields for both H and He are higher than the laboratory
results. In the proximity of the threshold, differences are larger.
In general, the threshold energies of the analytical yields are lower
than those derived with TRIM (with the exception of He into graphite). 
Such discrepancies may be the result of uncertainties in the estimate
of the energy parameters ($E_D$, $E_B$ and $E_S$) adopted in the TRIM
calculations, as discussed by \citet{FieldMNRAS1997}. In particular,
we found that a variation of 20 per cent in $E_S$ causes a variation of about
30 per cent in the sputtering yield at high energies, while the region of
the threshold depends more on the choice of $E_D$: by varying $E_D$ of
20 per cent the threshold energy changes by 25 per cent We stress again the 
uncertainty in the derivation of the yield in the proximity of the threshold:
the cross sections used by TRIM in computing the collisions at low energies
are based on extrapolations from experimental data which are only available at 
higher energies.

\begin{table*}
\centering
\begin{minipage}{180mm}
\caption{Best-fit parameters to be used in Eq.~\ref{eq_yfit} to calculate 
the angle-averaged sputtering yields of H and He on graphite and olivine. 
For olivine, parameters are given for each atomic species in the compound 
and for the total number of sputtered atoms. The parameters in the left panel 
refer to $\langle Y_i(E)\rangle_\theta$ as defined in Eq.~(\ref{eq_meany}) 
(the {\em streaming yield} adopted in this work). For completeness, in the 
right panel we also give the parameters derived using Eq.~(\ref{eq_meanymay}) 
to define the mean (the {\em isotropic yield}, preferred by
\citealt{FieldMNRAS1997} and \citealt{MayMNRAS2000}).}
\label{ta_paray}
\begin{tabular}{lllllcrrrrr}
\hline
\multicolumn{5}{c}{Streaming Yield}& &\multicolumn{5}{c}{Isotropic Yield}\\[3pt]
$k$  & $\beta$ & $\gamma$ & $E_\mathrm{th}$ & $E_\mathrm{max}$ && 
$k$  & $\beta$ & $\gamma$ & $E_\mathrm{th}$ & $E_\mathrm{max}$  \\[3pt]
     &  (eV)   &          &  (eV) &  (eV) &&      
     &  (eV)   &          &  (eV) &  (eV)  \\
\hline
\multicolumn{11}{c}{ H $\rightarrow$ Graphite } \\[4pt]
4.356E-2&3.997E+1&1.059E-1&8.133E+1&3.620E+2&
&6.322E-2&4.176E+1&1.165E-1&8.119E+1&1.049E+3
\\[8pt]
\multicolumn{11}{c}{ He $\rightarrow$ Graphite } \\[4pt]
3.301E-1&1.456E+1&1.249E-1&3.075E+1&1.095E+3&
&5.140E-1&1.518E+1&1.231E-1&3.072E+1&2.175E+3
\\[8pt]
\multicolumn{11}{c}{ H $\rightarrow$ Olivine } \\[4pt]
5.029E-3&1.486E+2&1.423E-1&2.011E+2&1.329E+3&
Fe 
&7.726E-3&1.480E+2&1.457E-1&2.020E+2&2.965E+3
\\
4.778E-3&1.386E+2&1.455E-1&2.033E+2&1.198E+3&
Mg 
&7.390E-3&1.544E+2&1.466E-1&2.007E+2&2.403E+3
\\
4.441E-3&1.078E+2&1.728E-1&2.094E+2&1.552E+3&
Si 
&7.020E-3&1.306E+2&1.569E-1&2.057E+2&2.739E+3
\\
1.789E-2&1.044E+2&1.588E-1&2.073E+2&1.469E+3&
O  
&2.842E-2&1.210E+2&1.477E-1&2.053E+2&2.537E+3
\\
3.207E-2&1.140E+2&1.584E-1&2.063E+2&1.444E+3&
Tot
&5.048E-2&1.301E+2&1.482E-1&2.042E+2&2.609E+3
\\[8pt]
\multicolumn{11}{c}{ He $\rightarrow$ Olivine } \\[4pt]
5.361E-2&5.502E+1&1.276E-1&7.056E+1&2.018E+3&
Fe 
&7.819E-2&5.993E+1&1.174E-1&6.994E+1&4.106E+3
\\
4.805E-2&6.232E+1&1.173E-1&6.962E+1&1.630E+3&
Mg 
&6.816E-2&7.180E+1&1.036E-1&6.851E+1&3.093E+3
\\
4.794E-2&5.780E+1&1.223E-1&7.056E+1&1.748E+3&
Si 
&6.991E-2&6.353E+1&1.166E-1&6.988E+1&3.336E+3
\\
1.893E-1&4.489E+1&1.294E-1&7.101E+1&1.873E+3&
O  
&2.741E-1&4.774E+1&1.193E-1&7.076E+1&3.373E+3
\\
3.388E-1&4.970E+1&1.276E-1&7.066E+1&1.863E+3&
Tot
&4.910E-1&5.316E+1&1.186E-1&7.028E+1&3.458E+3
\\
\hline
\end{tabular}
\end{minipage}
\end{table*}

The behavior of the sputtering yield for $\theta > 0^\circ$ depends
on the energy $E$. For high $E$ values, the yield increases with $\theta$
faster than the usually assumed $1/\cos\theta$ dependence, while at 
energies lower than the energy corresponding to the maximum yield, 
$Y_i(E,\theta)$ is flatter and decreases when the projectile approaches
grazing incidence. As a result, 
$\langle Y_i (E) \rangle_\theta > 2 Y_i (E,\theta= 0^\circ)$ at high
energies and $\langle Y_i (E) \rangle_\theta \approx Y_i (E,\theta=
0^\circ)$ close to the threshold
\citep[see also the discussion in ][]{JuracApJ1998}. 
TRIM calculations are made using a plane-parallel infinite target. 
Sputtering yields may increase for spherical grains of the dimension of
the mean penetration depth of a projectile \citep{JuracApJ1998}.
As we mainly deal with large grains, we do not consider this
radius-dependent enhancement.

The data points computed with TRIM were fitted with the following 
function:
\begin{equation}
Y(E)=k \exp\left[-\frac{\beta}{E-E_\mathrm{th}}-
\gamma\left(\ln\frac{E}{E_\mathrm{max}}\right)^2\right].
\label{eq_yfit}
\end{equation}
While the first term in Eqn.~(\ref{eq_yfit}) (introduced by 
\citealt{MayMNRAS2000}) well describes the rapid increase of the 
sputtering yield in the proximity of the threshold energy 
$E_\mathrm{th}$, the second term, centered on the energy value
$E_\mathrm{max}$, is needed to reproduce the maximum and the slow 
decrease of $Y(E)$ at higher energies. The function provides a
remarkably good description of both $Y_i(E,\theta)$ (see, for example 
the TRIM data points for $Y(E,\theta = 0^\circ)$ and their fit in 
Fig.~\ref{fi_yield0}) and the angle averaged yield 
$\langle Y_i (E) \rangle_\theta$ of Eqn.~(\ref{eq_meany}), with most 
of the data points ($\sim$ 95 per cent) being within less than 10 per
cent of the 
fitted function.  The best-fit values of $E_\mathrm{th}$, $E_\mathrm{max}$ 
and the parameters $k$, $\beta$ and $\gamma$ can be found in 
Table~\ref{ta_paray} (right panel)\footnote{The angle-averaged yield of 
Eq.~(\ref{eq_meany}) is referred to in \citet{FieldMNRAS1997} and
\citet{MayMNRAS2000} as the {\em streaming yield}, appropriate for 
projectiles which streams in one dimension and collides with a 
non-rotating spherical grain. However, for a isotropic distribution 
of projectiles it is more appropriate to use the {\em isotropic yield},
given by
\begin{equation}
\langle Y_i (E) \rangle_\theta=\int_{0}^{\pi/2} Y_i(E,\theta)
\sin\theta d\theta.
\label{eq_meanymay}
\end{equation}
While \citet{FieldMNRAS1997} and \citet{MayMNRAS2000} prefer the
latter, we use in this work the former. As they pointed out, differences
between the two means are not large. For completeness, we also give the 
best fit parameters for the {\em isotropic yield} in the right panel of
Table~\ref{ta_paray}.}. 
For olivine, sputtering yields are slightly different for each of the
elements in the compound. For simplicity, we use here the sputtering 
yield for the total number of atoms. Parameters for the fit of the total 
sputtering yield of olivine are also given in Table~\ref{ta_paray}.

\begin{figure*}
\epsfig{file=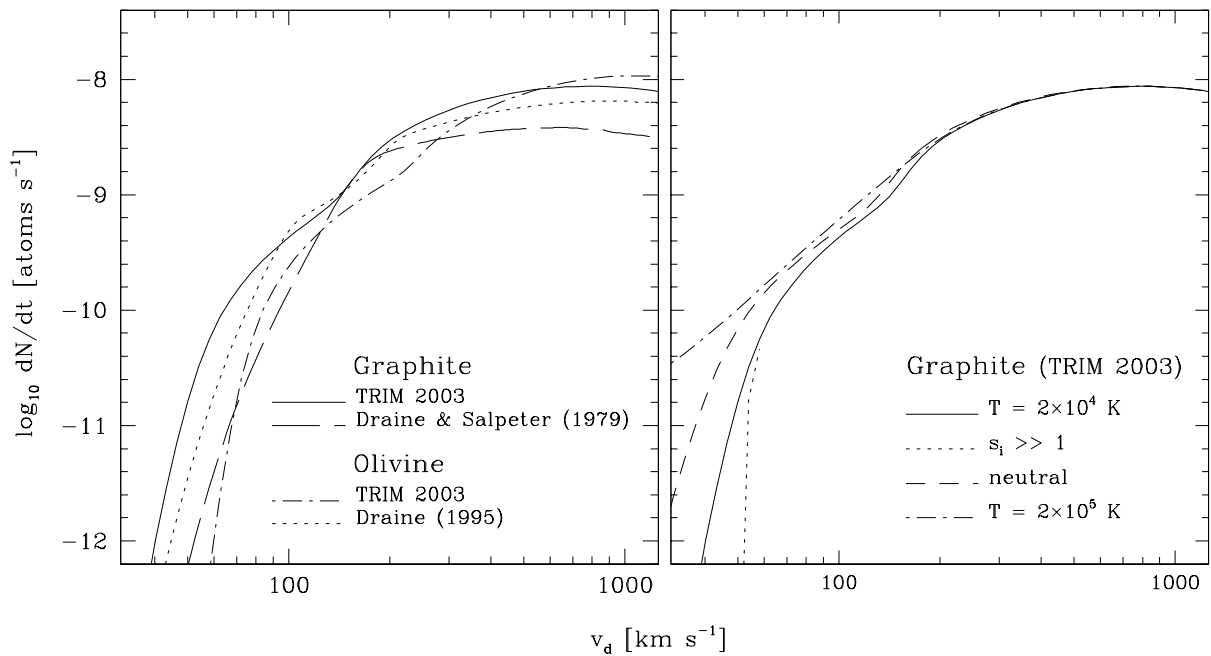,angle=0,width=16cm}
\caption{Sputtering rates for a graphite and olivine grain of radius $a$ = 
0.1-$\mu$m, moving with velocity $v$ through a gas with mean $z=3$ cosmic 
density. The gas is assumed to be composed of \ion{H}{ii}, \ion{He}{ii}
and electrons. In the left panel the sputtering rates are shown 
for the TRIM results (graphite- solid line, olivine- dotted-dashed line) and 
for the analytical model (graphite- dashed line, \citealt{DraineApJ1979a}; 
olivine- dotted line, \citealt{DraineApSS1995}), for a mean gas temperature 
$T=2\times 10^4$-K and a grain equilibrium potential $U$=20-Volts. In the 
right panel we show the TRIM 2003 sputtering rate for graphite
under different conditions. Solid line: same as in the left panel; dotted 
line: supersonic approximation (Eq.~\ref{eq_asputrate}); dashed line: neutral 
grain (or/and neutral gas); dotted-dashed line: $U$=20-Volts charged 
grain moving in a $T=2\times 10^5$-K gas.}
\label{fi_sputrate}
\end{figure*}

To show the differences between the analytical and the TRIM sputtering
yields, we plot in Fig.~\ref{fi_sputrate} (right panel) the sputtering 
rates computed from Eqn.~(\ref{eq_fsputrate}) for a graphite or olivine 
grain moving through a gas composed of \ion{H}{ii}, \ion{He}{ii} and
electrons. The grain has radius $a$ = 0.1-$\mu$m and
equilibrium potential $U=20$-eV. The gas has mean $z=3$ cosmic density and 
temperature $T=2\times 10^4$-K. For both analytical and TRIM sputtering 
yields, the low velocity sputtering rate is obviously dominated by collisions 
with the more massive He atoms. The contribution of the more abundant H atoms 
becomes important for $v> 100-200$ km s$^{-1}$ (where both He and H 
sputtering yields depend weakly on the impact energy). As said
for the normal incidence case, the largest differences between TRIM and 
analytical sputtering rates are for graphite grains. However, the difference
is small and the results shown in this paper do not change significantly 
when one of the yields is preferred over the other (see \S~\ref{se_homo}). 
For the mean conditions considered in this work, 
sputtering is mostly due to the grain motion ({\em non thermal} sputtering), 
as it can be seen in the right panel of Fig.~\ref{fi_sputrate}: for
$v\ga100$ km s$^{-1}$, the supersonic approximation (dotted line) is very 
close to the full calculation (solid line). In most cases, 
the grain potential acts to increase the effective sputtering 
threshold, because part of the impact energy has to be spent in overcoming 
the Coulomb repulsion between the positively charged grain and the ions 
acting as projectiles. This is shown by the higher sputtering rate for a 
neutral grain for $v\approx 100$ km s$^{-1}$ (left panel: dashed line). 
If the gas temperature increases (left panel: dot-dashed line)
the contribution of thermal sputtering becomes dominant for low velocities.

\subsection{Implementation}
\label{se_imp}

The motion of dust grains is studied in a three-dimensional grid (typically 
made of 128$^3$ cells). For each cell in the grid, gas density, temperature 
and ionization state are defined. After the determination of the position 
of the galaxies in the computational volume grid, grains are ejected into 
the IGM over the whole solid angle (from the center of each galaxy). The 
motion is assumed to start at a distance from the galaxy equal to 
its virial radius. A single ejected grain is thus characterised by its 
composition (graphite or olivine), radius $a$, velocity $v$ and path 
direction. All these initial quantities are derived  randomly with a Monte 
Carlo procedure from the adopted distribution (usually, those defined in
\S~\ref{se_eje}). Along the path the grain crosses cells with different 
gas conditions and its properties (velocity, charge and radius) change. 

As the grain enters a cell, it is instantaneously assigned an equilibrium
potential (the time necessary to reach the equilibrium, $t\approx
|Z/J_\mathrm{ce}|$ is usually much shorter than the time necessary for a 
grain to go across a cell; for the mean dust and gas conditions
considered here, $t\approx$25~yrs). For a fixed UVB, the equilibrium
potential will depend on the gas density, temperature and plasma
composition and on the grain material, radius and velocity. Given
the equilibrium potential and the other gas and dust properties, we can
compute the sputtering rate appropriate for the cell, $dN/dt$. The total 
number of atoms that are released as the grain moves through the cell is 
then $\Delta N$=$(dN/dt)(\Delta l/ v)$, where $\Delta l$ is the length of 
the grain path inside a cell. These sputtered atoms are deposited in the 
cell. As a result of sputtering, the grain loses mass (its radius reduces).

Finally, we compute the reduction in velocity, $\Delta v= -(\Delta l/ v)
F_\mathrm{drag}$, due to collisional and Coulomb drag. The grain velocity 
(and radius) is updated when the grain leaves a cell and enters the next 
along the path. The process is then repeated (computation of equilibrium 
charge, of deposited atoms and reduction in velocity) and the grain motion 
followed for the chosen time duration of the simulation $t_\mathrm{f}$
(unless the velocity is so small that the grain will not move more than one 
cell in the remaining time). If $\Delta v$ is large in a cell the grain path 
is split in smaller units and $\Delta v$, charge and $\Delta N$ are computed
for each of them. We found that assuming $|\Delta v|/v<0.1$ for each path 
segments provides accurate results without overly increasing the computational
time. For the cosmological simulation of \S~\ref{se_cosmo}, path splitting is 
typically required at the end of a grain's path, for velocities below 100 km 
s$^{-1}$ (the simulations of \S~\ref{se_homo} are computed with a larger
spatial resolution).

In practice, the Monte Carlo procedure is run for $N_\mathrm{p}$ cycles
(or grain {\em packets}) to assure a good statistics (i.e.\
sampling of initial grain properties and path directions). If
$N_\mathrm{g}$ is the total number of grains that are ejected into the
IGM in a simulation (defined by the mass of ejected dust and the size 
distribution), in each cycle $N_\mathrm{c}=N_\mathrm{g}/N_\mathrm{p}$ 
identical grains will be ejected along a chosen direction. Thus, for each 
cell, $N_\mathrm{c}\times \Delta N$ atoms are deposited in the IGM. 
The final output of the simulation is a grid with the number density of 
sputtered atoms. The final positions of the dust grains along their path 
is also stored.

\section{Results: homogeneous density}
\label{se_homo}

\begin{figure*}
\epsfig{file=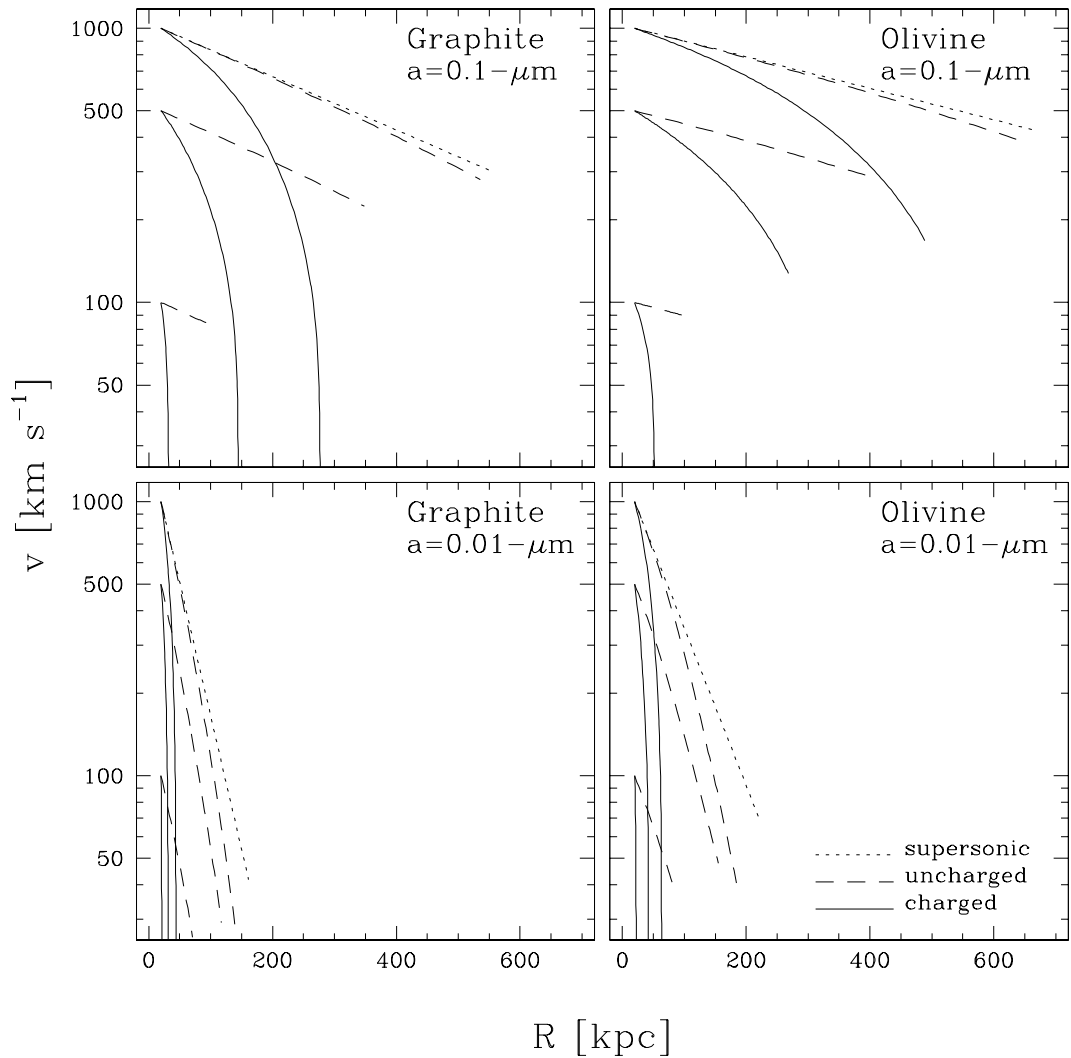,angle=0,width=14truecm}
\caption{
Velocity of dust grains injected into a homogeneous medium with 
mean $z=3$ cosmic density, as a function of the distance from the source
galaxy, $R$. For each panel, we show the velocity change of three grains
with initial velocities (at $R_\mathrm{V} = 20$ kpc) $v_0 = $ 100, 500
and 1000 km s$^{-1}$ (from bottom to top). The two upper panels show the 
velocity of graphite (left) and olivine (right) grains of radius 
$a$=0.1-$\mu$m. The two lower panels show the same, but for
$a$=0.01-$\mu$m. Solid lines show the results when the whole
physics described in \S~\ref{se_physics} is taken into account, while 
dashed lines refer to the case in which the grain charge has been set to 
zero. For the $v_0 = $ 1000 km s$^{-1}$ we also show the solution for
the neutral supersonic approximation of grains with a fixed radius
(Eqn.~\ref{eq_vsupa}). The gas is made of \ion{H}{ii}, \ion{He}{ii} and 
electrons and has T=$2 \times 10^4$-K. 
The TRIM 2003 sputtering yields have been used (\S~\ref{se_sput}).
}
\label{fi_veloh}
\end{figure*}

To understand the details of grains motion and sputtering, we first analyze 
the case of a single galaxy ejecting grains into a homogeneous medium with z=3 
mean cosmic density. The IGM is assumed to be composed of \ion{H}{ii}, \ion{He}{ii} 
and electrons. Unless it is said otherwise, the gas temperature is assumed to be 
$T=2\times 10^4$-K and the TRIM 2003 sputtering yield are used in the calculations.
As we said in \S~\ref{se_imp}, we study the motion of the grain from the virial 
radius $R_\mathrm{V}$, defined as the radius of the sphere enclosing a mean density 
of 200 times the critical density \citep{NavarroApJ1997}. In the cosmological 
simulation described in the next section, a galaxy of {\em median} mass has a 
dark matter halo of 1.5$\times$10$^{10}$ M$_\odot$ (and a baryonic content of 
1.5$\times$10$^{9}$ M$_\odot$); the appropriate value for the virial radius 
is $R_\mathrm{V}$ = 20 kpc (proper units). We follow the motion of each grain for 
$t_\mathrm{f}$=1 Gyr: the results of the simulation can thus be compared to 
the observed properties of the universe at $z=2$. During this time stretch, 
the UVB flux increases only by 25 per cent \citep{BianchiA&A2001}; we neglect 
this time dependency in the calculations. 

It is instructive to study the velocity evolution of dust grains of
different size and material. In Fig.~\ref{fi_veloh} we show $v$ as 
a function of the distance $R$ from the center of the galaxy.
Grains are ejected at $R_V$ with three different initial velocities 
$v_0= 100, 500$ and $1000$ km sec$^{-1}$. 
The two top panels refer to the case of grains of radius $a$=0.1-$\mu$m,
while in the two bottom panels $a$=0.01-$\mu$m. Panels to the left are for
graphite grains, those on the right for olivine. With the dashed line we 
show the results obtained for a grain whose charge is kept neutral (i.e.\ 
only the collisional part is included in the drag force). For the same
initial velocity, larger grains can attain larger distances from the 
galaxy: despite having larger cross sections when colliding with the IGM
atoms, their deceleration is smaller because they are heavier. For the
same reason, heavier olivine grains travels to larger distances R. 
This dependence on the grain size and grain material density can be 
derived analitycally for a neutral grain of radius $a$ moving at 
supersonic speed. By integrating Eqn.~\ref{eq_coll_s} and \ref{eq_asputrate}
and neglecting the erosion of the grain due to sputtering (the radius $a$ 
is kept constant), the velocity of the grain as a function of the distance 
$R$ from the center of the galaxy can be written as
\begin{equation}
v(R>R_\mathrm{V})=v_0 \exp\left[ -\frac{3(R-R_\mathrm{V})}{4 a \rho} \Sigma_i n_i m_i\right].
\label{eq_vsupa}
\end{equation}
The supersonic approximation for neutral grains of Eqn.~\ref{eq_vsupa} 
is shown as a dotted line in Fig.~\ref{fi_veloh} (for ease of presentation,
only for $v_0$=1000 km sec$^{-1}$).  The difference between the
approximation and the curve for the case with only collisional drag is due 
to grain erosion. Despite small grains have smaller sputtering rates
(see Eqn.~\ref{eq_fsputrate}) they contain less
atoms and their size reduces more. For $v_0$=1000 km sec$^{-1}$, 
$a$=0.01-$\mu$m grains lose in the IGM approximately $3\times 10^5$ atoms, 
reducing their radius by 30-40 per cent  (graphite and olivine, respectively). 
Conversely, $a$=0.1-$\mu$m grains lose $2\times 10^8$ atoms but their 
radius reduces less, by 20-25 per cent.

The results for the complete calculation including grain charge and
Coulomb drag is shown in Fig.~\ref{fi_veloh} by the solid lines.
When exposed to the $z$=3 UVB, 0.1-$\mu$m graphite grains attain a
potential of about 25-Volts, while the potential of olivine grains is 
about 20-Volts (the exact value depending on the grain velocity;
\S~\ref{se_coc}). Because of Coulomb drag, the maximum distance $R$
to which a grain can travel in the time $t_\mathrm{f}$ is reduced.
Having a smaller charge (together with the larger mass), high velocity
olivine grains can still travel at considerable distances (400-500 kpc)
from the galaxy with velocities (kinetic energies) above the threshold
for significant sputtering. Instead, high velocity graphite grains 
eventually stop at $R\la$300 kpc. Grains with $a$=0.01-$\mu$m attains
potentials a couple of Volts larger than those with $a$=0.1-$\mu$m. 
As even in the neutral case small grains were not able to go beyond 
$R\approx$200 kpc, when the Coulomb drag is included they stop 
within 50 kpc from the ejection, even for the less restrictive
case of a $v_0$=1000 km s$^{-1}$ olivine grain. Small grains do not
travel out to significant distances in the IGM. An analogous result 
has been obtained for the dust ejection from galaxies by some of the 
authors listed in \S~\ref{se_eje}, although with a more detailed study 
of the grain motion, including gravitation and radiation pressure.
This is why from now on we will use, as described in \S~\ref{se_eje}, 
a flat size distribution, with 0.05$<a $[$\mu$m] $<$0.2, suggested by 
literature models. Together with this we will assume a flat distribution 
of initial velocities, within the range 100$<v<$1000 km s$^{-1}$.

\begin{figure*}
\epsfig{file=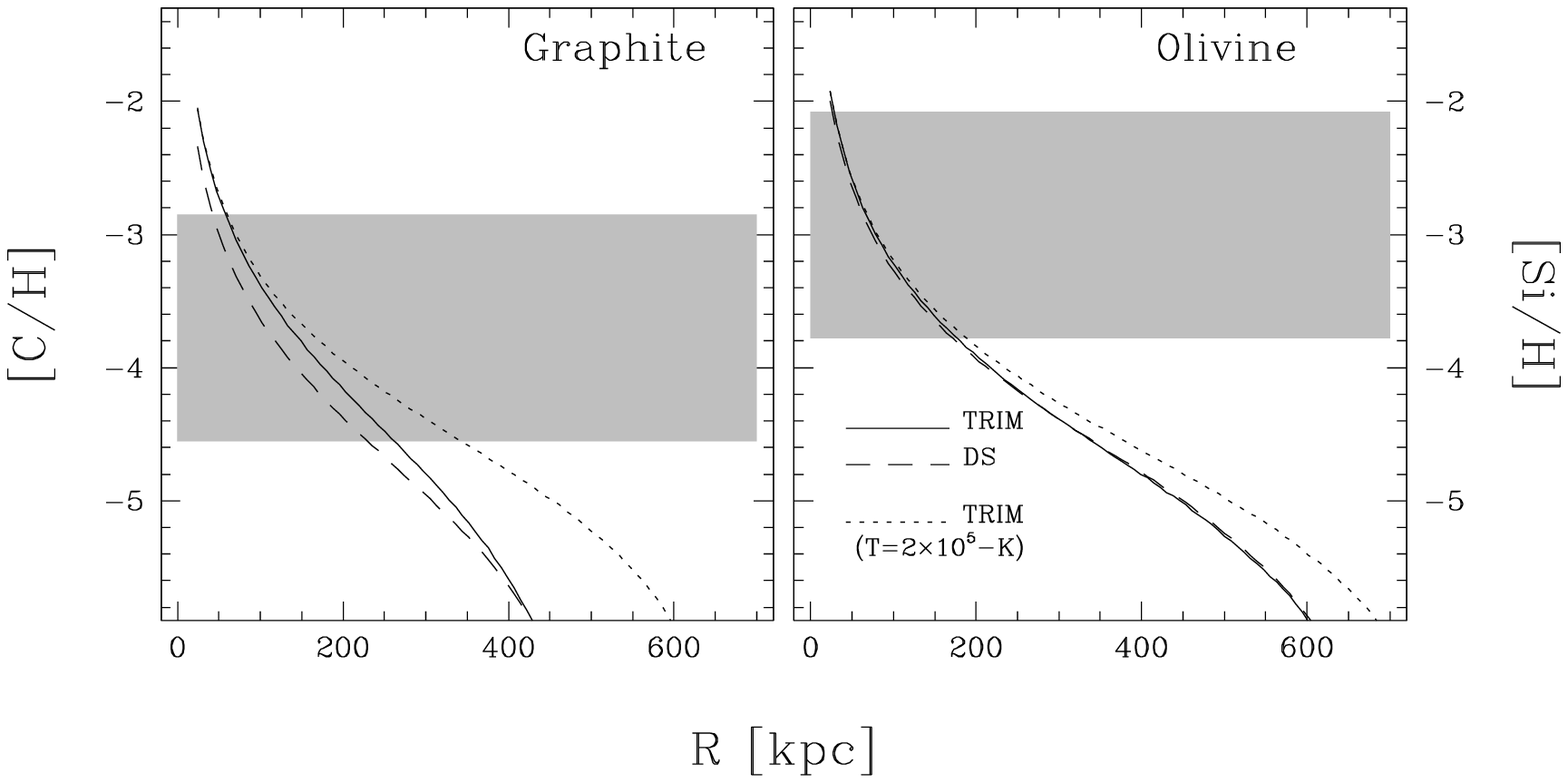,angle=0,width=14truecm}
\caption{
IGM metal pollution due to dust grains injected into a homogeneous medium 
with mean $z=3$ cosmic density, as a function of the distance from the 
source galaxy, $R$. The left panels show the carbon metallicity [C/H]
resulting from erosion of graphite grains, while the right panels
show the silicon metallicity [Si/H] resulting from erosion of olivine grains.
For each material, a flat distribution in the range 0.05$< a$ [$\mu$m] 
$<$ 0.2 has been used for the grain radii and a flat distribution
in the range 100 $< v_0$[km~s$^{-1}$] $<$ 1000 for the grain
initial velocities. The gas is made of \ion{H}{ii}, \ion{He}{ii} and 
electrons and has T=$2 \times 10^4$-K. In total, $3\times10^5$ M$_\odot$ 
of dust has been ejected. 
The results shown with solid lines have been obtained using the TRIM
2003 sputtering yields, while for the dashed lines the Draine \&
Salpeter's yields have been used. The dotted lines refer to the TRIM
2003 case, but the gas temperature has been raised to $2\times 10^5$-K.
The gray areas show the metallicities derived at 
$z=2$ for the overdensity $\delta = 1$ \citep{SchayeApJ2003,AguirreApJ2004}.
}
\label{fi_metadi}
\end{figure*}

Fig.~\ref{fi_metadi} displays the carbon and silicon IGM
metallicities\footnote{The metallicity of element X is given by 
$[\mathrm{X}/\mathrm{H}] = \log(n_\mathrm{X}/n_\mathrm{H}) -
\log(n_\mathrm{X}/n_\mathrm{H})_\odot$, where $n_\mathrm{X}$ is
the number density of element X and $(\mathrm{X}/\mathrm{H})_\odot =
\log(n_\mathrm{X}/n_\mathrm{H})_\odot$ is the solar abundance. 
We use $(\mathrm{C}/\mathrm{H})_\odot= -3.44$ and
$(\mathrm{Si}/\mathrm{H})_\odot= -4.45$ \citep{AndersGeCoA1989}.
}
resulting from the sputtering of carbon and 
silicon\footnote{As described in \S~\ref{se_sput}, we have used the
yield for the total number of atoms sputtered off olivine. The number 
density of Si (and Fe, Mg) is thus 1/7 the total number density of 
atoms released by olivine grains in the IGM (the O number density 
being four times higher). The use of the rescaled total sputtering 
yield  provides a good approximation to the specific sputtering 
yields for each of the species composing the material.} 
atoms off graphite (left panel) and olivine grains (right panel), 
for the adopted size and velocity distributions. The metallicity is 
reported as a function of the distance from the galaxy, $R$.
The metallicity levels produced in our simulations depend linearly 
on the number of grains $N_\mathrm{g}$ ejected (equivalent to the 
ejected dust mass, for a fixed size distribution). The mass of dust
ejected in Fig.~\ref{fi_metadi} has been derived from the baryonic 
mass of our {\em median} $z=3$ galaxy, by assuming that, at maximum,
1/500 of the baryonic mass is in dust in a galaxy \citep{EdmundsMNRAS1998}
and that 10 per cent of the total dust mass can reach the IGM (see 
Sect.~\ref{se_eje}). In total, 3$\times$10$^{5}$ M$_\odot$ of dust 
(both graphite and silicate) are ejected into the IGM for the 
simulation of Fig.~\ref{fi_metadi}. Though we believe this to be a 
reasonable upper limit, the metallicity results can be easily scaled 
for any desired ejected mass of dust.

We compare the results of our simulation with the metallicities of C
and Si derived by \citet{SchayeApJ2003} and \citet{AguirreApJ2004}. They
analyzed the pixel optical depth in a set of high resolution observations 
of the Ly$\alpha$ forest using numerically simulated spectra and
estimated the metallicity as a function of redshift and gas density. 
The gray areas in Fig.~\ref{fi_metadi} show the median metallicity ($\pm$
1$\sigma$ of lognormal scatter) for gas of mean density (overdensity 
$\delta$=1) at $z=2$. The [Si/H] value has been extrapolated from the ratio
[Si/C] \citep{AguirreApJ2004}, with the caveat that it has been derived
only for gas with $\delta > 3$. The scatter plotted for [Si/H] is that 
derived for [C/H]. \citet{SchayeApJ2003} find no significant trend for 
[C/H] versus redshift in the range $1.8 < z < 4.1$. A detailed study
of the metal enrichment history is needed to asses whether this lack of
evolution can be explained by dust sputtering or galactic winds
\citep{AguirreApJ2001a}, or if instead an early enrichment from pre-galactic
($z\approx 9$) sources is necessary \citep{MadauApJ2001}. Here, we 
simply want to check if it is possible to reproduce, by using the dust 
sputtering mechanism only, the same level for [C/H] and [Si/H] as 
inferred from observations.

Fig.~\ref{fi_metadi} shows that it is possible to obtain metallicity
levels similar to those inferred from observations, at least for
distances from the galaxy of $R\la200$ kpc. Within this range,
the metallicity levels are quite similar for both materials.
As we saw in Fig.~\ref{fi_veloh} olivine grains can travel to larger 
distances and pollute with metals a larger area. However, the difference 
in the trend can be appreciated only at $R\ga300$ kpc, where grains 
have reduced velocities and thus produce low levels of metallicity.
We remind the reader that we have assumed, as an upper limit, the 
same relative abundance of graphite and silicate grains in the 
ejected dust. If silicates have a smaller ejection efficiency, as 
some studies suggest (see Sect.~\ref{se_eje}) the metallicity level 
in Fig.~\ref{fi_metadi} will be lower (but the spatial distribution 
will be the same).

A change in the gas temperature does not affect significantly the
results. The dotted curve shows the case for $T$=$2\times 10^{5}$-K. 
The metallicity becomes higher only for larger $R$, due to grains that
are not supersonic anymore (see, e.g., the analogous case in 
Fig.~\ref{fi_sputrate}). Again, since graphite grains are slowed down
more than those made of olivine, the difference is higher for [C/H]
than for [Si/H].  Because the sputtering yield is quite flat for
high energies, a further increase in the gas temperature does not produce
a dramatic change in the results. Even for an extreme case in which
$T$=$2\times 10^{7}$-K, the metallicities are only about 0.3-0.4 dex higher
than those for $T$=$2\times 10^{4}$-K.
In Fig.~\ref{fi_metadi} we also show the results when the traditional
Draine \& Salpeter sputtering yields are used (dashed lines). The
differences are small, especially when compared to the large
uncertainties and scatter of the metallicities inferred from
observations. Graphite grains produce a lower [C/H]  (but by only 0.2
dex) when using Draine \& Salpeter yields, while the difference is 
smaller for olivine. This is mainly due by the difference in sputtering
rates for $v>$200 km s$^{-1}$, which is larger for graphite than for
olivine (see the left panel of Fig.~\ref{fi_sputrate}). For the rest of
the paper, we will use the TRIM 2003 sputtering yields.

\section{Results: Cosmological simulation}
\label{se_cosmo}

\begin{figure*}
\epsfig{file=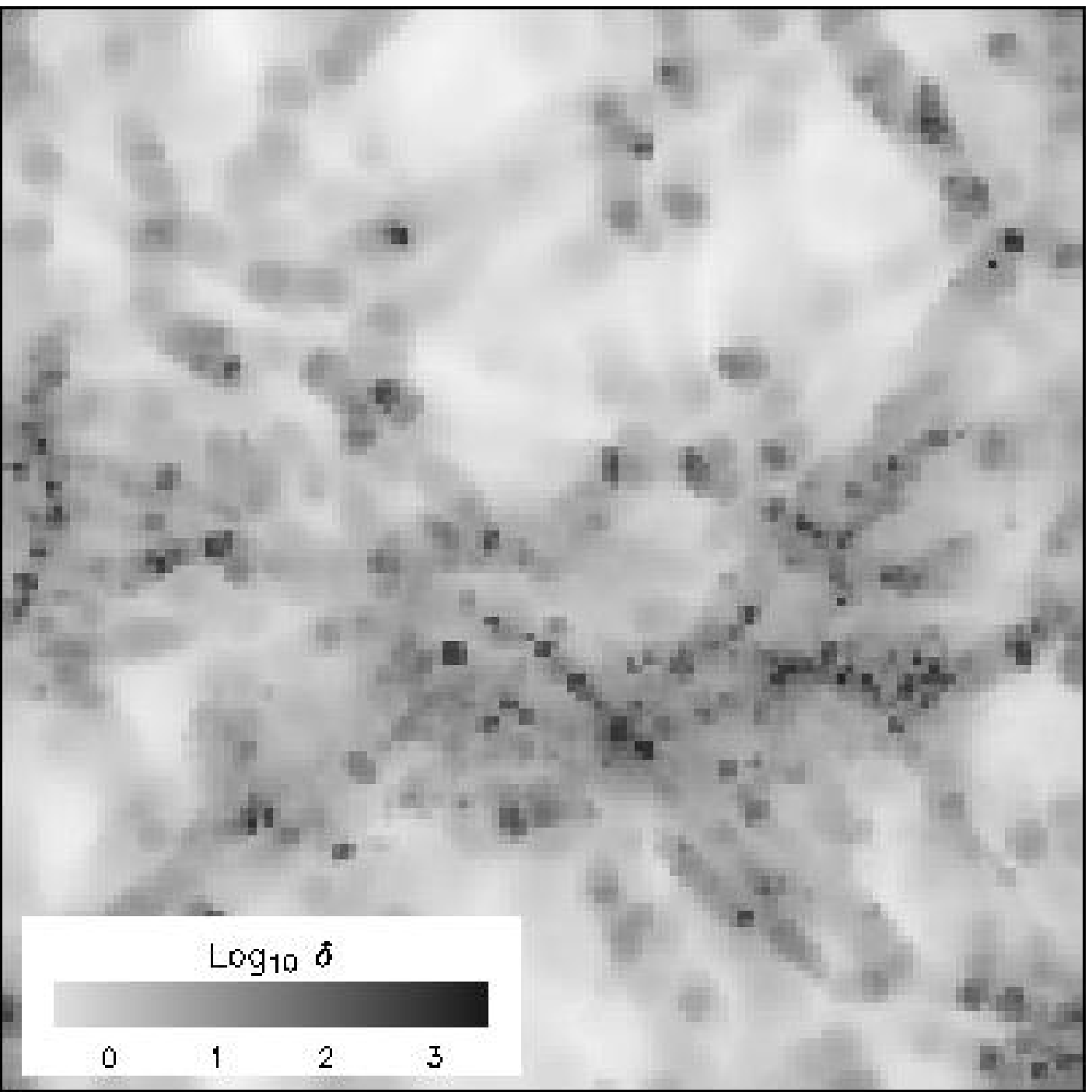,angle=0,width=5.8truecm}
\epsfig{file=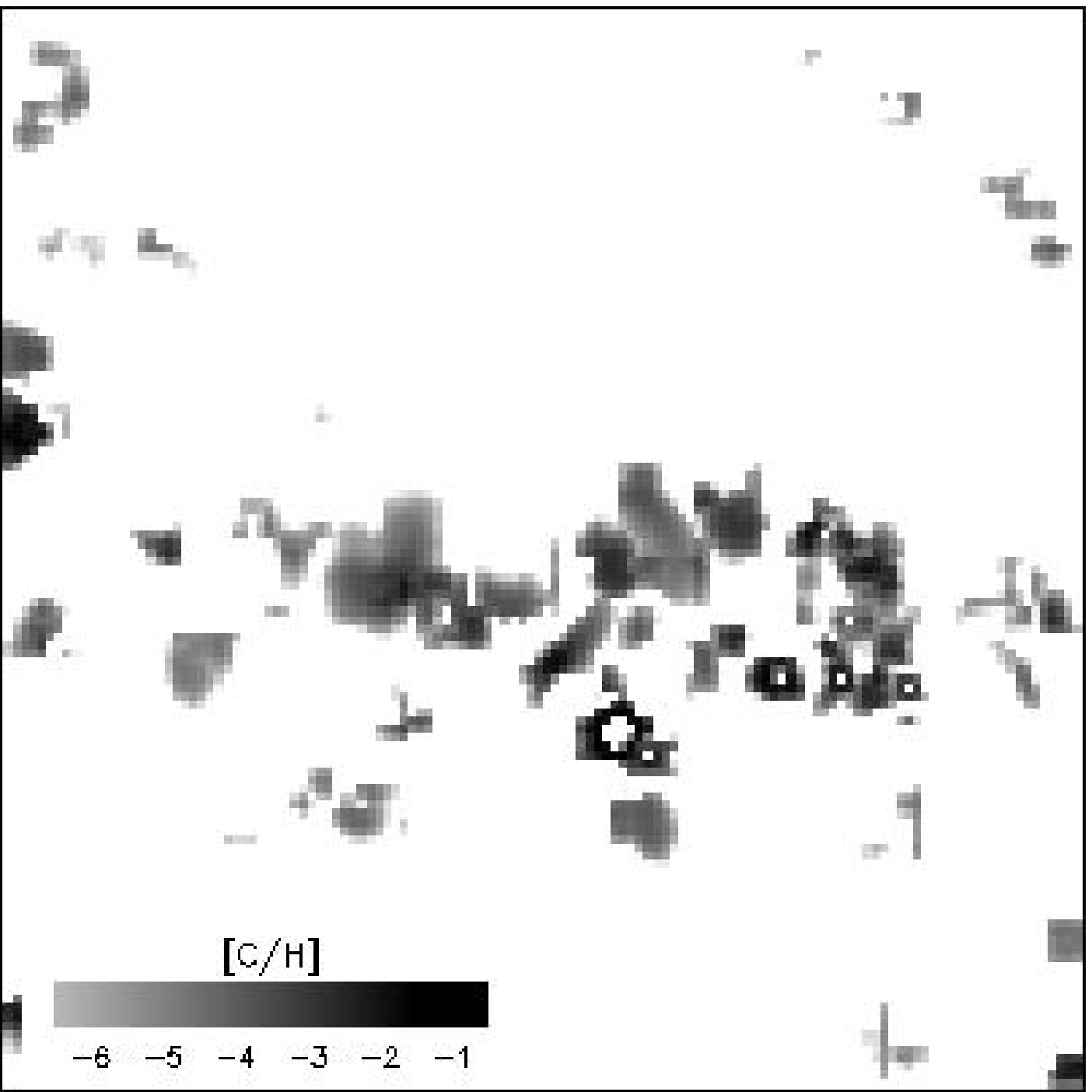,angle=0,width=5.8truecm}
\epsfig{file=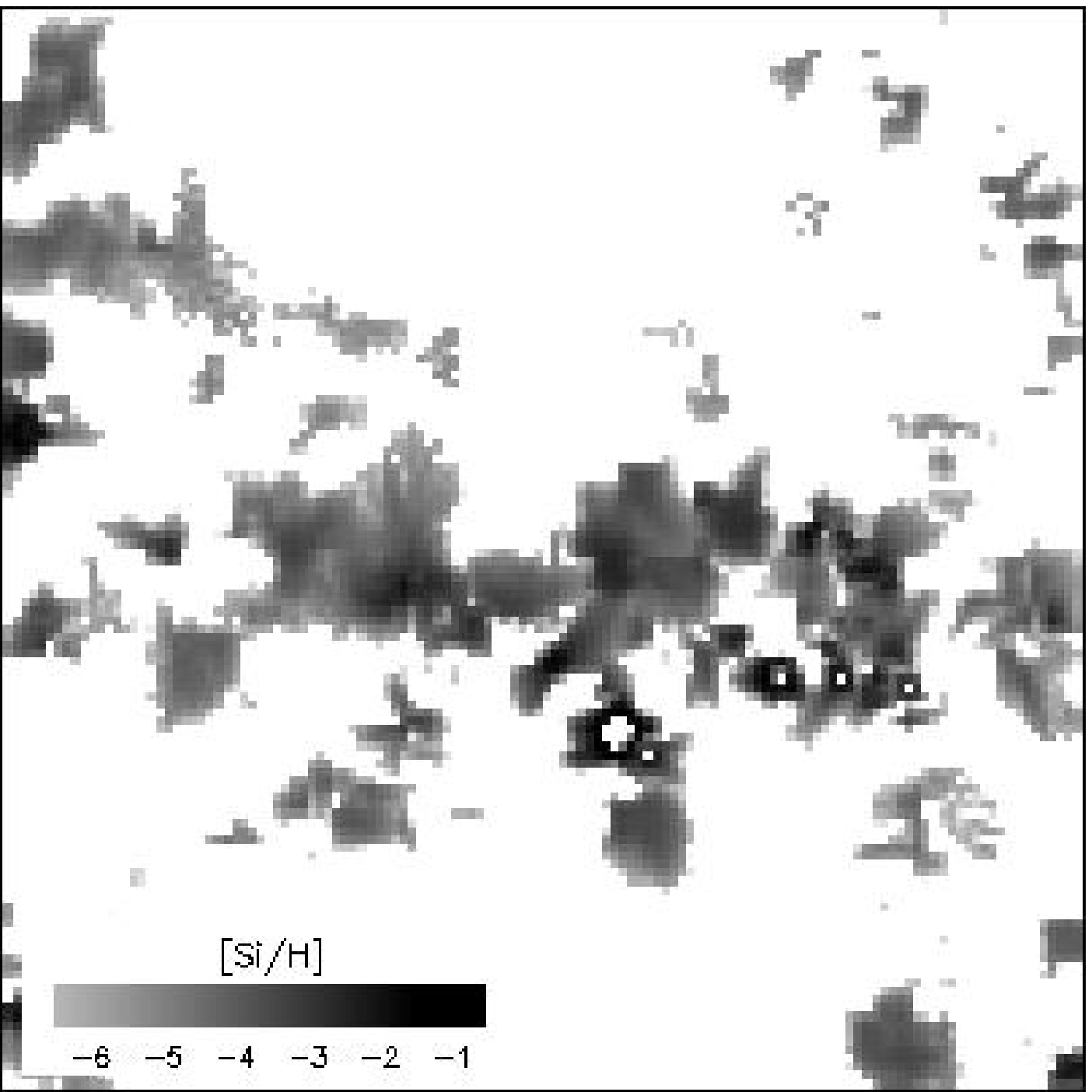,angle=0,width=5.8truecm}
\caption{
Cuts through the simulation box. The left panel shows the gas overdensity,
while the centre and right panels show the carbon and silicon metallicities,
respectively. The side of each map is 10.5$h^{-1}$ comoving Mpc, 3.5 physical 
Mpc at $z=3.27$. The unpolluted (white) cells in the [C/H] and [Si/H] maps
are internal to the virial radius of the most massive galaxies in the 
selected slice.
}
\label{fi_maps}
\end{figure*}

We now study the motion of grains in a cosmological density field.
The simulation has been obtained using a multi-phase SPH code particularly 
designed for the study of galaxy formation \citep{MarriMNRAS2003} and 
galactic winds. 
The initial parameters are those for the $\Lambda$CDM model adopted so far,
($\Omega_0=0.3$, $\Omega_\Lambda=0.7$, $\Omega_b h^2=0.028$, $\sigma_8=0.9$).
The simulation uses $128^3$ particles in a box of 10.5$h^{-1}$ comoving Mpc. 
A group finding algorithm has been used to identify dark matter halos and the 
star-forming galaxies associated to them. For a more detailed description of 
the simulation, we refer the reader to \citet{BruscoliMNRAS2003} 
and \citet{MaselliMNRAS2004}.

\begin{figure*}
\epsfig{file=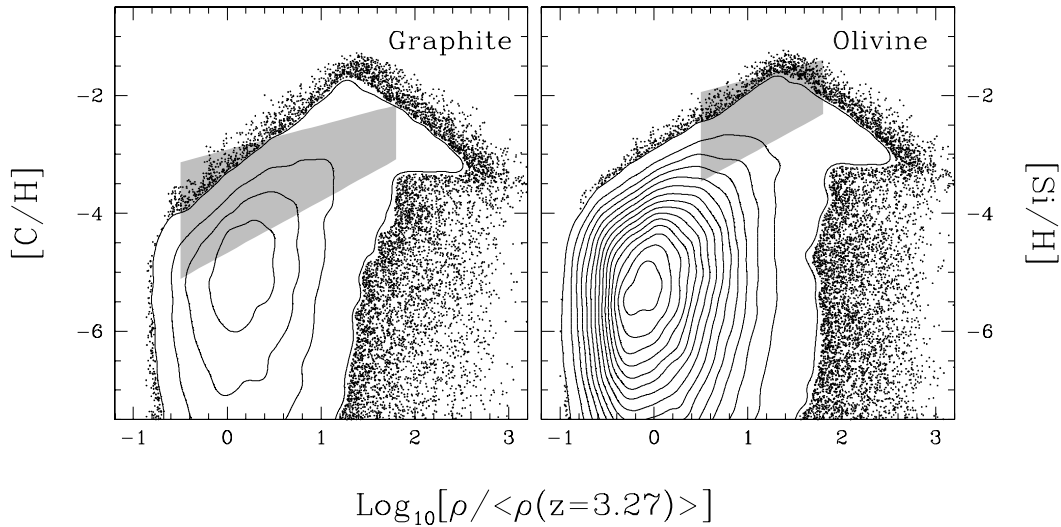,angle=0,width=14truecm}
\caption{
Metallicity vs overdensity for simulation cells polluted by carbon
from graphite grains (left) and silicon from olivine grains (right).
Each cell is indicated by a dot. Contours are plotted for regions
where the density of dots becomes high. In both panels, contours
start at 20 (and increase by 40) cells per 0.1 dex in overdensity 
and metallicity. The gray area refers to the metallicity measurements of
\citet{SchayeApJ2003} and \citet{AguirreApJ2004}.
}
\label{fi_metovd}
\end{figure*}

\begin{figure}
\center{
\epsfig{file=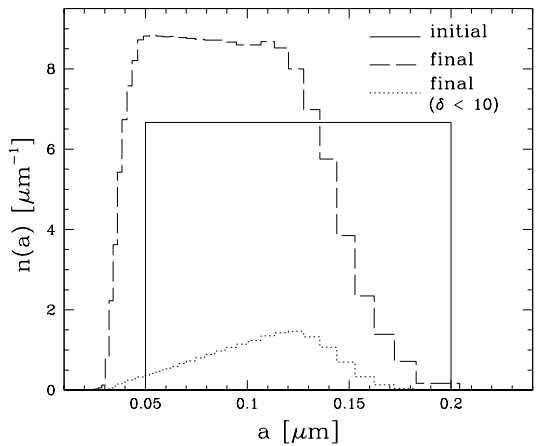,angle=0,width=7truecm}
}
\caption{
Initial (solid line) and final (dashed line) grain size distributions.
The two histograms have been normalised to unity. The distribution for
grains at rest in gas with $\delta < 10$ is shown with a dotted line 
(about 10 per cent of the total).
}
\label{fi_gsd}
\end{figure}

\begin{figure}
\center{
\epsfig{file=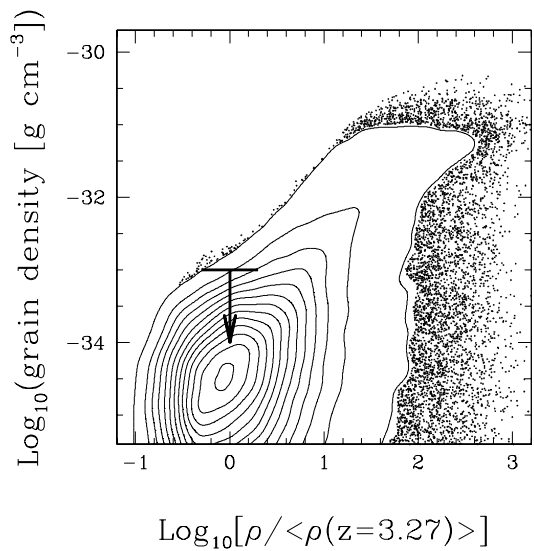,angle=0,width=7truecm}
}
\caption{
Dust grain density in the IGM vs gas overdensity for simulation 
cells occupied by grains at rest. Each cell is indicated by a dot. 
Contours are plotted for regions where the density of dots becomes 
high; they start at 20 (and increase by 80) cells per 0.1 dex in 
overdensity and grain density. The upper limit refers to the results 
of \citet{InoueMNRAS2004}.
}
\label{fi_gdeovd}
\end{figure}

We investigated the sputtering process on the $z=3.27$ simulation
output. The particle properties have been mapped on a 3-D grid of 
128$^3$ cells. For each cell, we derived gas density, temperature
and ionization fractions for each of the chemical species by 
performing a standard SPH smoothing on the 32 particles nearest to 
each cell. The spatial resolution of the grid is $\approx$82 kpc 
$h^{-1}$ comoving ($\approx$27.5 kpc physical). In total, 398 groups 
of particles have been identified as galaxies in the simulation box. 
The mass of dust in each galaxy has been derived from the baryonic 
mass, using the dust-to-baryonic mass ratio computed by 
\citet{EdmundsMNRAS1998}. Again, we allow 10 per cent of the dust 
to be ejected from the virial 
radius\footnote{In total, 2.7$\times 10^8 M_\odot$ of dust are 
injected into the simulation volume, corresponding, for the adopted 
dust distributions, to a total number of dust grains $N_\mathrm{g}=
1.6\times 10^{55}$. We have run simulations with $N_\mathrm{p}=10^9$ 
packets of 1.6$\times 10^{46}$ grains each. This is sufficient for
the results presented in this section to be independent of
$N_\mathrm{p}$ (i.e. Fig.~\ref{fi_maps} to \ref{fi_metovd_f2} are not 
affected by simulation cells where only a few grain has passed and thus
not having converged to a mean value for the metallicity, grain
density and number).}
As both the dust-to-baryonic mass ratio and the fraction of dust
ejected are considered to be upper limit, we believe the metallicity 
results obtained in this paper (which scale with the number of ejected 
grains) to be upper limits to the dust contribution to the IGM pollution.
The virial radius is computed from the mass of the halo associated to 
each galaxy, as specified in \S~\ref{se_homo}. Less massive galaxies 
are obviously more represented, with 80 per cent of the sample having a 
baryonic mass between $8.5\times 10^8$ M$_\odot$ (the minimum mass)
and $3.0\times 10^9$ M$_\odot$ (and a dark matter mass roughly 10 
times that value). As we said in \S~\ref{se_homo}, a galaxy of 
{\em median} mass ejects into the IGM $3\times 10^5$ M$_\odot$ of dust. 
Smaller objects also have virial radii which are quite close to the cell 
resolution (the halo of the smallest object has roughly the volume 
of a single cell).

In Fig.~\ref{fi_maps} we show a cut through the whole simulation box,
parallel to one of the box faces. The map in the left panel refers
to the gas density field; the metallicity maps in the central and
right panel ([C/H] and [Si/H], respectively) are the results of the
calculations described in this paper. As discussed previously, the drag
has a smaller effect on the heavier and less charged silicate grains.
This is evident in the larger area of the slice polluted with silicon.
The slice passes through the most massive galaxy in the simulation
(with baryonic mass $7\times 10^{11}$ M$_\odot$) which can be
identified in the metallicity maps with the largest "hole" (inside a 
high metallicity region). The holes are the regions within the halo
of each galaxy, that are excluded from the simulation (as grains are
assumed to move from the virial radius {\em outward}). For the largest
galaxy, $R_\mathrm{V}$=70 kpc physical (2.5 times the dimension of a 
cell). Other high mass objects can be seen to the right. 
It is interesting to note 
how the massive galaxies are not necessarily surrounded by a wider
metal enriched area, despite injecting into the IGM a larger amount 
of dust. As they reside in denser gas, grains ejected from these
objects are more easily stopped by the drag.

In Fig.~\ref{fi_metovd}, we show the metallicity as a function of the
gas overdensity. Each dot in the left (right) panel represents a simulation
cell which has been polluted by carbon (silicon) as the result of the
passage of one or more graphite (silicate) grains. As a reference, in the
left panel we have plotted the median carbon metallicity ($\pm$
1$\sigma$ of lognormal scatter) derived by \citet{SchayeApJ2003} as a
function of the gas overdensity (gray area). We are showing the value at 
$z=2$. In the right panel, the same area is indicated, but scaled for
the [Si/C] ratio measured by \citet{AguirreApJ2004} for gas with
$\delta>3$. Similar metallicity levels have been reported by
\citet{SimcoeApJ2004}, by fitting hydrogen and metal lines at $z=2-2.5$
in QSO's absorption spectra. They do not detect a significant
trend with the gas density. However, this may be due to the different
choice for the UV background needed to correct for ionization, as harder 
spectra lead to a shallower dependence on $\delta$ \citep{SchayeApJ2003}.

As seen in the homogeneous case, grain sputtering can produce non
negligible metallicities in the IGM gas. The dependence of metallicity 
on density is steeper than what inferred from observations. Most cells 
with $\delta\approx1$ have [C, Si/H] $\approx$ -5, approximately one 
order of magnitude lower than the values measured by \citet{SchayeApJ2003} 
and \citet{AguirreApJ2004} (especially [C/H]). 
The contribution of dust sputtering becomes more important for moderately
overdense gas with $\delta=10-100$, which can be translated in the neutral 
hydrogen column density range $14.5 < \log N(\ion{H}{i}) < 16$. 
We find, however, that only 4-5 per cent of the cells in this overdensity
range have [Si,C/H] values within the observationally allowed area.
We remind that we believe this to be an upper limit, both for the total 
amount of dust and for the relative number of silicate and graphite  
grains ejected into the IGM (we use a 1:1 ratio). If the fraction of silicate 
grains is smaller, as some works seem to suggest (\S~\ref{se_eje}), [Si/H] 
would be smaller. However, the trend with density will be the same.
While the level of metallicity in our simulations scales with the ejected 
mass of dust, the value derived by \citet{SchayeApJ2003} depends on their 
assumption for the UV background. The scatter, instead, depends less on
the assumptions and more on the underlying physics and on the density
structure of the IGM (Aguirre, private communication). For cells with 
$\delta\approx 1$, we find a log-normal scatter of about 1.6 dex (slightly 
larger for Si). This is about twice what inferred from observations.

The features in the scatterplot reflect the movement of dust grains.
In each panel, the region in the top-right is occupied by cells 
{\em just outside} the virial radius of the injecting galaxies, 
where the grain movement starts. In principle, 
for an isolated halo, gas at $R_\mathrm{V}$ has $\delta\approx$60 
\citep{NavarroApJ1997}. However, due to the complex IGM structure and 
to the lack of resolution, the gas cells where the grain movement starts
have overdensities $1\la\log_{10}(\delta)\la3$ (we will comment on 
resolution in \S~\ref{se_confro}). Because of the high density and grain 
velocities, metallicities  (and sputtering rates) are high in these
cells. As the grain moves towards the less dense gas, 
it is slowed down by gas drag. For reduced velocities
and gas densities, the sputtering rates are lower and so the
resulting metallicities. This explains the general trend from
high $\rho$ - high metallicity to low $\rho$ - low metallicity seen
in both panels of Fig.~\ref{fi_metovd}. The slope of the relation
depends both on the initial grain velocity and radius (as seen for the
homogeneous case of \S~\ref{se_homo}) and on the gas density.

There are some exceptions to such trend. Grains ejected into the higher 
density gas are stopped by drag more efficiently and pollute a smaller
region around them. This is why the high metallicity cells at $\delta>2$ 
appear isolated (i.e.\ with no contiguous lower metallicity and $\delta$ 
points) in the scatterplot. As the velocity quickly reduces within the 
dense region, some high-$\delta$ cells end up with low metal pollution. 
In this case the slope of the relation becomes very steep and lack of 
resolution limits the detailed study.
Other low metallicity cells with high-$\delta$ mark the end-point of 
grains previously travelling in a less dense environment that stop
as they cross a denser filament in the density field.

The gas which grains encounter as they leave a galaxy is often shock-heated by
gravitational collapse and/or SNe. The cells with the highest metallicities in 
Fig.~\ref{fi_metovd} have temperatures in excess of 10$^5$-K. However,
this has little effect on the metallicity levels: the grain initial
velocities considered here are always higher or close to those 
delimiting the non-thermal regime for sputtering. Instead, the high 
temperature determines the mechanism responsible for the drag.
In the hot gas, collisional drag is more important (\S~\ref{se_drag})
and the grain movement does not depend on the charge. Coulomb drag becomes
more important than collisional drag in the low temperature gas close to
$\delta$=1 and it is because of it that grains come to a halt. Within the
adopted simulation time $t_\mathrm{f}$, most grains are slowed down 
and have energies below the threshold energy for sputtering. Even a ten-fold 
increase in $t_\mathrm{f}$ does not change the results significantly.

By counting cells which have been polluted, we can provide an estimate for 
the volume filling factors of metals. A wider volume is polluted by
silicon, and this is clearly shown by the larger number of points (cells) 
in the right panel of Fig.~\ref{fi_metovd} (and by the map in 
Fig.~\ref{fi_maps}). By checking simulations with increasing number of grain
packets, we found that the silicon filling factor tends asymptotically to about 
18 per cent. As most of the cells are occupied by low density gas far from sources, 
filling factors are obviously higher for denser regions. When considering
all cells with overdensity $\delta > 10$, the silicon filling factor 
increases to 40 per cent. The mean filling factor of cells polluted with carbon
is a factor three lower ($\approx$6 per cent). Thus, the distribution of metals 
in our simulations is rather inhomogeneous. Though a fraction of cells
in Fig.~\ref{fi_metovd} have metallicities compatible with the median 
value of \citet{SchayeApJ2003}, the median metallicity in our simulation
is zero, for any gas density bin.

As Fig.~\ref{fi_metovd} clearly show, the ratio [Si/C] in our simulation
is lower than $0.77\pm0.05$, the value inferred by \citet{AguirreApJ2004} 
at $z\approx3$ for gas with $\delta \ga 3$. In the high density gas, [Si/C] 
depends mostly on the assumption about the relative number of silicate/graphite 
grains and on the differences between the sputtering yields. If a cell is 
crossed by the same amount of both kind of grains, and if the kinetic energy 
is above the sputtering threshold, the ratio is quite constant. For example, 
silicate and graphite grains travelling at 500 km s$^{-1}$ in the same cell 
would pollute the gas to [Si/C]$\approx$0.15. Indeed, the high density cells
where the grain movement starts have [Si/C]$\approx 0.1\pm 0.08$. Again,
a comparison between the scatter in simulations and that derived from
observations may be less dependent on the assumptions that have been
made: for the high density cells, the scatter is similar to that of
\citet{AguirreApJ2004}, while it increases at lower density. The [Si/C] value 
increases as well at lower $\delta$, because olivine grains keep higher
velocities to larger distances (for our assumptions, 3 per cent of the 
simulated volume has [Si/C]$>1$). 

During their travel, dust grains are eroded but never destroyed by sputtering. 
At the end of the simulation, dust grains at rest are present throughout all
the volume polluted with metals (the filling factors of metal polluted cells
and cells containing grains at rest are similar). The final size distribution 
(Fig.~\ref{fi_gsd}) is roughly flat, 
as the input one, but with a different range in radii. The smallest grains have 
radii $a\approx$0.02-$\mu$m and originate from particles with initial radius 
$a$=0.05-$\mu$m, while largest grains reduce their radius from $a$=0.2-$\mu$m 
to $a$=0.15-$\mu$m\footnote{Incidentally, by using TRIM \citet{GrayMNRAS2004} 
have found that sputtering does not simply lead to grain erosion: the stopping 
of projectiles into the material may change the structure and composition of 
the grain, especially in a metal rich gas}.
The latter are the grains that contribute mostly to metal 
pollution and, as we discussed previously, travel farther out from their 
injection point: for regions of diffuse gas ($\delta < 10$) the grain 
distribution is peaked around sizes 0.1-0.15 $\mu$m. There is no significant 
difference between graphite and silicate grains.

In Fig.~\ref{fi_gdeovd}, we show the final grain density (both silicate and
graphite) as a function of the gas overdensity, for cells occupied by
dust grains at rest. Features corresponding to those of Fig.~\ref{fi_metovd} 
can be noticed. The higher grain densities, corresponding to higher gas 
densities, are due to a relatively large number of grains that stop 
close to the injection point. Only a limited number of grains travel far out, 
and this makes for the reduction of grain density for $\delta < 10$. 

\citet{InoueMNRAS2004} derived an upper limit for the dust grain density.
After adopting various cosmic star formation histories (to account for
the dust generation) they constrained the amount of dust in the IGM at
$\delta = 1$ with the maximum extinction and reddening allowed by the (now 
preferred) cosmological explanation for the dimming of high-z Type Ia SNe. 
Furthermore, they tightened the constraint requiring that dust grains cannot 
significantly alter, via photoelectric heating, the thermal history of the 
IGM derived from the low column density lines of the Ly$\alpha$ forest 
\citep{SchayeMNRAS2000}. The upper limit for grains of size $a=0.1$-$\mu$m 
is plot in Fig.~\ref{fi_gdeovd}. The median density for $\delta=1$ cells
occupied by grains is about 1.5 orders of magnitude lower than the upper 
limit. As for metallicity, the results of Fig.~\ref{fi_gdeovd} scale 
linearly with the amount of dust ejected in the IGM. If our upper limits
are correct, the extinction and heating effects of IGM grains are not 
likely to be detected.

\section{Comparison with previous works}
\label{se_confro}

\citet{AguirreApJL2001,AguirreApJ2001a} found that dust expulsion and
erosion can account for the observed levels of carbon and silicon enrichment 
in the IGM. Since there are several differences between their work and ours,
it is not easy to compare the results. They position grains at a
distance from the galaxy where gravitation is balanced by radiation
pressure. Because of this, silicate grains (having a larger material density)
do not lay as far out as graphite grains. Furthermore, silicate grains have 
a smaller efficiency for radiation pressure. This is opposite to our results,
where heavier silicate grains can travel to larger distances. They adopt
essentially a power law distribution for grain sizes. Grains with radii
$a=0.03-0.05 \mu$m have larger equilibrium distances than grains with
$a>0.1$-$\mu$m. Again, this is opposite to what we find here by studying the 
dynamics of grains.

Another difference is in the amount of dust that is ejected.
\citeauthor{AguirreApJL2001} allow half of the metal content of each
galaxy in their simulation to be distributed in the IGM as dust. As
typically only half of a galaxy's metal mass is believed to be in dust,
this is equivalent to ejecting the whole dust content from the galaxy. 
Although we believe that 10 per cent is an upper limit to the fraction
of dust ejected into the IGM (see \S~\ref{se_eje}), even allowing all
the dust mass to travel to the IGM (with large grains overrepresented
with respect to the standard MRN power law) does not change
significantly the results. As it can be seen by simply shifting upward
by 1 dex the metallicities of Fig.~\ref{fi_metovd}, more metal polluted 
cells will
lay within the range derived from observations. Still, the filling
factor of metal polluted cells will not change. The same shift applies 
to the points of Fig.~\ref{fi_gdeovd}. Even increasing by one order of 
magnitude the grain density, the limit set by \citet{InoueMNRAS2004} 
will not be violated in most cells with grains at rest.

\begin{figure*}
\epsfig{file=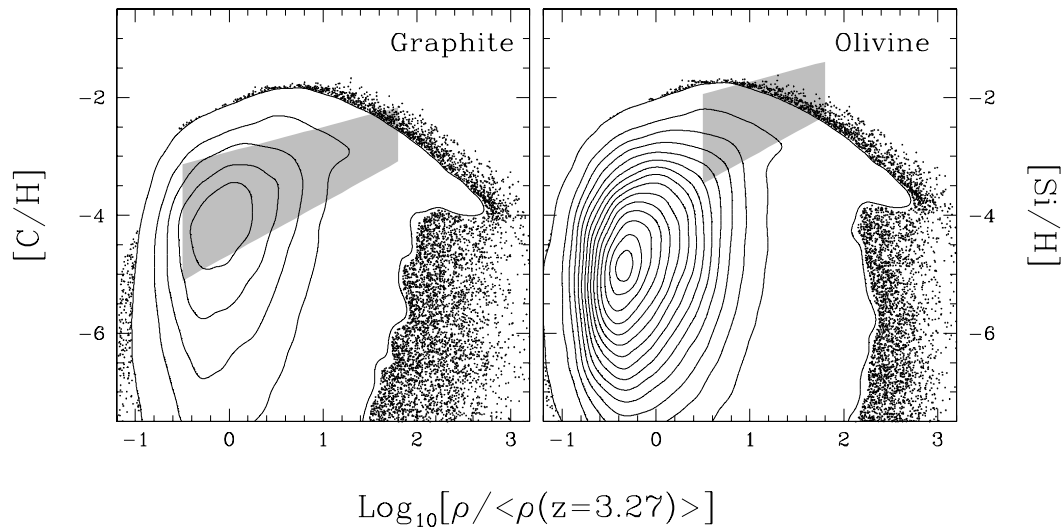,angle=0,width=14truecm}
\caption{
Same as Fig.~\ref{fi_metovd}, but ejecting dust grains at a distance
from each galaxy {\em twice} the virial radius. In these figure, 
contours start at 20 (and increase by 160) cells per 0.1 dex in 
overdensity and metallicity.
}
\label{fi_metovd_f2}
\end{figure*}

The distribution of metals in our simulations is inhomogeneous, with
filling factors smaller than 0.5 for any gas density bin, and it seems
smaller than that in \citet{AguirreApJL2001,AguirreApJ2001a} (we compare
our results to their case in which only the metals effectively eroded
from grains are considered). One potential interpretation is that in our
simulation the IGM density field has been ``frozen'' at $z=3$, where all 
grains have been ejected. The 40 per cent decrease of the mean density 
from $z=3$ to $z=2$ (the end of the simulation) it is however unlikely to 
produce significantly different effects. 
The lowest mass galaxies in our simulations have halos of the same size as
the cells. Lacking higher spatial (and density) resolution, grains
ejected from small galaxies at the virial radius may find themselves in
a denser environment than expected. As both these issues concern the
movement of grains in less dense environments, we ran a simulation in
which grains were injected at a distance from each galaxy {\em twice} the
virial radius. The metallicity-density plot for this test simulation is 
shown in Fig.~\ref{fi_metovd_f2}. Now the cells where the grain movement 
starts have a slightly lower density (there are less atoms 
to collide with) and obviously have lower metallicity (the peak of its 
distribution is shifted towards $\delta\approx$1). The reduction in gas 
particles also mean a smaller drag and a smaller reduction in velocity as 
the grain moves. The slope of the metallicity-density trend in
Fig.~\ref{fi_metovd_f2} becomes flatter, and
closer to the data of \citet{SchayeApJ2003}. Also, a good fraction of
$\delta\approx1$ cells has now metallicities within the
shaded area (at least for [C/H]). There is also an obvious increase in
the filling factors, with silicon being present in 50 per cent of the volume, 
and carbon in 20 per cent. Similar trends can be seen in the grain density 
distribution, with an increase of the density by about 0.2 dex for cells
with $\delta<1$ (a reduction by 0.2 dex for $\delta>10$). Clearly,
a better spatial resolution is desirable, although the conclusions are not 
going to be substantially altered. 

Finally, we note here that it would be possible to obtain higher 
metallicity levels by destroying completely the grains, via 
some unknown mechanism other from thermal/non-thermal sputtering. 
In our simulation a sizable fraction of large grains are at rest in 
gas with $\delta < 10$. By adding their contribution in atoms to the 
metallicity produced by the sputtering processes we have considered,
we would rise by almost 2 dex the metallicity of one third of cells 
at $\delta < 1$, while leaving metallicity levels unaltered for 
$\delta > 10$. 

\section{Summary and conclusions}
\label{se_summa}

We have analysed the motion of dust grains in the $z=3$ IGM, to study
the erosion due to sputtering and the subsequent pollution of the gas 
with metals. A fraction of the dust grains in a galaxy are thought to
escape the gravitational well because of radiation pressure from
starlight. Using results in the literature, we have defined plausible
distributions for the sizes and velocities of the escaping grains.
Our fiducial grain size distribution is flat, with radii in the range 
$0.05<a$[$\mu$m]$<0.2$, as literature results suggest that only large 
grains can escape from galaxies. The assumed velocity distribution is
flat as well, with $100<v$[km s$^{-1}$]$<1000$. 
For such velocities, grains are supersonic under most of the gas conditions 
explored: thus, {\it non-thermal} sputtering is the dominant mechanism for 
the grain erosion.

The motion of each ejected grain is studied from the virial radius,
until the velocity becomes small and the grain stops. Our computation
takes into account:
\begin{enumerate}
\item a calculation of the grain charge, due to collisions with
electrons and ions in the IGM plasma and to photoejection of electrons
by a metagalactic UV background. We have used the \citet{BianchiA&A2001}
UV background including the contribution of galaxies and taken into
account the velocity dependence of the collisional charging rates;
\item the gas drag, including both the collisional term and the Coulomb
term due to interaction of the (heavily) charged grain in the $z=3$ gas
and the charged particles in the gas;
\item the sputtering of atoms off the grain surface, due to collisions
with the ions in the IGM. Using the code TRIM, we have derived new
sputtering yields for H and He atoms colliding with graphite and a
silicate, olivine. Analytical fits to these yields are provided. The new
yield have been compared to the widely used yields of
\citet{DraineApJ1979a}.
\end{enumerate}

The grain motion has been studied for a single source of grains (galaxy)
in a homogeneous density field and inside a cosmological simulation,
allowing the galaxies detected in the simulation box to eject 10 per
cent of their dust mass into the IGM. We believe this assumption (as
well as the resulting metallicities and grain density) to be an upper 
limit, compatible with the presumable MRN-like dust distribution inside a
galaxy and the adopted size distribution for ejected grains.
The motion is followed for 1 Gyr (half the Hubble time at $z=3$), allowing 
us to compare our results to observations at $z=2$. The main results are:
\renewcommand{\labelenumi}{(\arabic{enumi})}
\begin{enumerate}
\item Only large ($a\ga 0.1$-$\mu$m) grains can reach out for the low
density regions in the IGM, because the drag is less effective for
heavier grains. For the same reason, silicate grains are able to diffuse
in the IGM more than graphite grains and pollute with metals a larger 
fraction of the cosmic volume. Carbon and silicon metallicities 
produced by dust sputtering of graphite and olivine grains show a well 
defined trend with gas density.  High density cells in the vicinity of 
galaxy halos (where the grains start their travel) have high metallicities, 
whereas fewer grains (originally at high $v$) reach gas with 
$\delta\approx$1, with reduced velocity producing smaller sputtering rates. 
Those grains that are able to reach lower density regions are stopped, within 
the simulation time, by Coulomb drag. None of the large grains is completely 
destroyed by sputtering. The results summarised in this point are general and
do not depend on the assumption on the adopted size and velocity distribution 
and fraction of dust mass ejected from galaxies, nor on the specific density 
field used in this work.

\item 
For our parameter choice, most of the $\delta\approx1$ simulation cells polluted
with metals have [C, Si/H] $\approx$ -5, approximately one order of magnitude lower 
than the median values recently derived by \citet{SchayeApJ2003} and 
\citet{AguirreApJ2004}.  The contribution of dust sputtering becomes more 
important for moderately overdense gas with $\delta=10-100$, corresponding 
to neutral hydrogen column densities $14.5 < \log N(\ion{H}{i}) < 16$. For 
this overdensity range, however, only 4 per cent of the cells have [C/H] 
values within the observationally allowed area. For any density bin, 
metallicities have a large scatter (a lognormal $\sigma > 0.15$ dex at 
$\delta\approx$1). This is twice what has been inferred from observations 
of absorption lines.
In addition, we recall that olivine grains pollute the IGM with Fe, Mg 
(same level as [Si/H]), and oxygen ([O/H] is a factor four higher than 
[Si/H]), although these species are less readily observed in the IGM.

\item Polluted cells in the range where dust sputtering is important 
($10 < \delta < 100$) tend to have 0$\la$[Si/C]$\la$0.2, almost 
independently of the gas density. The value we have obtained strongly
depend on the graphite/silicate proportion, which we have assumed to be 
1:1, as for MRN dust. If this is correct, we would have [Si/C]$\la$0.2
when allowing for a lower ejection efficiency of silicate grains with
respect to graphite (as suggested by a few work in the literature).
Such ratio is substantially smaller than the nucleosynthetic one, 
approximately equal to 0.9. Hence, this ratio might represent a valuable 
tool to assess the importance of the grain sputtering enrichment by 
systematic searches and studies of low [Si/C] absorption systems.

\item Resulting metal distributions are very inhomogeneous, with only
18 per cent of the volume occupied by cells polluted with silicon (from 
far-reaching olivine grains) and 6 per cent with carbon. These results 
slightly depend on resolution, but we estimate that filling factors larger 
than 50 per cent(for silicon) for cells at $\delta=1$ enriched by dust 
sputtering are unlikely. 

\item Even allowing all dust mass produced inside galaxies by $z=3$ to be 
ejected, the grain density in the low ($\delta \approx 1$) density IGM is 
below the upper limit derived by \citet{InoueMNRAS2004}, from IGM thermal 
history and extinction considerations. However, some moderately dense 
regions with $\delta > 10$ might contain a significant amount of dust, 
whose extinction effects remain to be calculated in detail.
\end{enumerate}

To conclude, our results show that radiation-pressure driven dusty flows, 
followed by sputtering of the grains, might represent a viable and attractive 
mechanism to enrich the IGM with (cool) heavy elements. This process 
mostly affects moderately overdense IGM, without affecting too much the 
low-density gas: this is understood by the enhanced sputtering efficiency 
where the gas density is higher. 

At least, three aspects of the problem need further study before we can draw 
final conclusions. The first concerns the ejection of dust grains from
galaxies: although a few authors have studied the problem (we have
used their results here as initial conditions) so far no work presents a 
detailed statistical analysis of the properties of escaping grains. In
particular, estimates are lacking for the fraction of the dust mass
which is ejected, as a function of the dust size and composition and of
the galactic mass. This is important if we want to study the history of
dust ejection and metal pollution by grain sputtering. At reddhift
$z>3$, despite an higher IGM density (increase in the drag), the physical 
distance between galaxy is smaller and grains may be able to pollute a
larger volume of the universe. However, galaxies have a smaller stellar
mass: if the internal radiation field is weaker, the dust ejection 
efficiency may be reduced (unless this effect is contrasted by changes
in the ISM drag and in gravity). 
The second aspect relates to the UV background fluctuations: in this 
study we have considered the UVB to be isotropic. However, radiative transfer 
effects (as shadowing and shielding), in the vicinity of galaxies or dense 
cosmic filaments, might create anisotropies resulting in a net radiation 
force on the grain. Such force might be large enough to prevent the grain 
from stopping as it currently happens in our simulations. In this case, the 
grain can continue to move essentially taking a random walk, until it enters 
a regions of very high density (for example an accreting flow) within which 
it becomes fully coupled to the gas.
If so, more metals could be released and less grains would survive, thus 
further decreasing the resulting intergalactic extinction.
The third important effect concerns the intergalactic magnetic fields, which 
we have so far ignored.  Although they might not be generally dynamically 
important, they might affect dust sputtering, via the betatron effect. 
We plan to focus on these additional physical effects in forthcoming papers. 

\section*{Acknowledgments}
SB would like to thank the European Southern Observatory, where this 
project started, and the Osservatorio Astrofisico di Arcetri (OAA), 
for hospitality during the early phases of the work. AF acknowledges 
the affiliation to the OAA. We are grateful to A. Inoue and A. Aguirre
for profitable discussions, to L. Del Zanna and S. Landi for their help
with computational issues, and to an anonymous referee for comments that 
significantly improved the paper. 
This work was partially supported by the Research and 
Training Network "The Physics of the Intergalactic Medium" set up by 
the European Community under the contract HPRN-CT2000-00126 RG29185.

\bibliography{../../DUST}

\label{lastpage}
\end{document}